\documentclass[aps,pra,twocolumn,a4paper,showpacs]{revtex4}
\usepackage{amsmath}
\usepackage{amssymb}
\usepackage{version}
\usepackage{graphicx}
\usepackage{dcolumn}
\usepackage{psfrag}

\newcommand{\beq}{\begin{equation}}
\newcommand{\eeq}{\end{equation}}
\newcommand{\cond}{\Phi_0 (\mathbf{r})}
\newcommand{\ccond}{\Phi_0^* (\mathbf{r})}

\newcommand{\hsp}{\hat{H}_{\text{sp}}}
\newcommand{\nt}{\tilde{n}}
\newcommand{\mt}{\tilde{m}}
\newcommand{\bfr}{\mathbf{r}}
\newcommand{\bfrp}{\mathbf{r'}}
\newcommand{\rt}{(\mathbf{r},t)}

\newcommand{\rrw}{(\mathbf{r},\mathbf{r'},\omega)}

\newcommand{\rw}{(\mathbf{r},\omega)}

\newcommand{\intdr}{\int \!\! d \mathbf{r} \,}

\newcommand{\bpm}{\begin{pmatrix}}
\newcommand{\epm}{\end{pmatrix}}
\newcommand{\percent}{\%}

\newcommand{\col}[2]{\begin{pmatrix} #1 (\mathbf{r}) \\ #2 (\mathbf{r}) \end{pmatrix}}
\newcommand{\coldgPw}{\begin{pmatrix} \delta \Phi \rw \\ \delta \Phi^{*} (\mathbf{r},-\omega) \end{pmatrix}}

\newcommand{\bfs}{\mathbf{s}}
\newcommand{\bfq}{\mathbf{q}}
\newcommand{\bfp}{\mathbf{p}}

\newcommand{\epr}{\epsilon(\mathbf{p},\mathbf{r})}

\begin{document}

\title{Quantitative test of thermal field theory for Bose-Einstein condensates II}
\author{S. A. Morgan}
\email[]{sam@theory.phys.ucl.ac.uk}
\affiliation{Department of Physics and Astronomy, University College London, Gower Street, London WC1E 6BT, UK}

\date{\today}

\begin{abstract}
We have recently derived a gapless theory of the linear response of a Bose-condensed gas to external perturbations at finite temperature and used it to explain quantitatively the measurements of condensate excitations and decay rates made at JILA [D. S. Jin \textit{et al.}, Phys. Rev. Lett. {\bf 78}, 764 (1997)]. The theory describes the dynamic coupling between the condensate and non-condensate via a full quasiparticle description of the time-dependent normal and anomalous averages and includes all Beliaev and Landau processes. In this paper we provide a full discussion of the numerical calculations and a detailed analysis of the theoretical results in the context of the JILA experiment. We provide unambiguous proof that the dipole modes are obtained accurately within our calculations and present quantitative results for the relative phase of the oscillations of the condensed and uncondensed atom clouds. One of the main difficulties in the implementation of the theory is obtaining results which are not sensitive to basis cutoff effects and we have therefore developed a novel asymmetric summation method which solves this problem and dramatically improves the numerical convergence. This new technique should make the implementation of the theory and its possible future extensions feasible for a wide range of condensate populations and trap geometries.
\end{abstract}


\maketitle


\section{Introduction} \label{INTRODUCTION}

Measurements of the excitation spectrum of Bose-Einstein condensates (BECs) provide a unique opportunity to compare quantitative predictions of finite temperature quantum field theories (QFTs) with experimental data in a regime where interactions, finite temperature and particle statistics are all important. At low-temperature the experimental results can be understood using only the Gross-Pitaevskii equation (GPE) \cite{GPE} for the condensate, but the situation is less satisfactory at finite temperature where a description of the coupled dynamics of condensed and uncondensed atoms is required. There are also more fundamental difficulties with developing a consistent finite temperature theory, such as dealing correctly with the ultra-violet and infrared divergences which can appear and obtaining a gapless spectrum, as required by the Hugenholtz-Pines theorem \cite{Hugenholtz59}. We have recently developed a gapless theory which addresses all these issues and provides a general method for calculating the response of a BEC to external perturbations at finite temperature \cite{Morgan04}. We have demonstrated the validity of this approach by applying it successfully to the pioneering measurements of condensate excitations made at the Joint Institute for Laboratory Astrophysics (JILA)  in 1997 \cite{Jin97,Morgan03a}. In particular, we were able to explain the sudden upwards shift in the resonance frequency of the low-lying $m=0$ mode as the temperature increased towards the critical temperature for BEC formation. In this paper we describe our numerical calculation and provide a detailed analysis of the results and their implications in the context of this experiment. The theoretical approach and numerical techniques are general, however, and can be applied directly to a wide variety of recent experiments on BECs \cite{StamperKurn98,Marago01,Chevy02,Bretin03}. 

The early experiments at JILA in 1996 and 1997 \cite{Jin96,Jin97} measured the energies and decay rates of low-lying condensate excitations with axial angular momentum quantum numbers $m=2$ and $m=0$ as a function of the condensate population and temperature. The low-temperature measurements are in good agreement with the usual Bogoliubov quasiparticle theory \cite{Jin96,Stringari96,Edwards96} and with calculations based on the Hartree-Fock-Bogoliubov (HFB) formalism \cite{Griffin96}, but the finite-temperature results have proved much harder to explain. In the experiment of \cite{Jin97}, the $m=2$ mode was observed to shift strongly downwards with temperature, while the $m=0$ mode underwent a sharp increase in energy towards the non-interacting limit. The results for the $m=2$ mode could be explained using a gapless extension of the HFB approach (GHFB) which includes the so-called anomalous (pair) average of two condensate atoms \cite{Proukakis98,Hutchinson98}, and also by the dielectric formalism \cite{Reidl00}. However, both approaches were unable to explain the upward shift of the $m=0$ mode, and an analytical calculation also predicted downward shifts for both modes \cite{Giorgini00}.

A possible explanation for the behaviour of the $m=0$ mode was given by Bijlsma, Al Khawaja and Stoof \cite{Bijlsma99,Khawaja00} in terms of a crossover from out-of-phase to in-phase oscillations of the condensed and un-condensed atoms at high temperature. Subsequently, Jackson and Zaremba \cite{Jackson02} obtained good agreement with the JILA results for both modes using the Zaremba-Nikuni-Griffin formalism \cite{Zaremba98,Nikuni99,Zaremba99} which involves coupling a generalized GPE for the condensate to a a semi-classical Boltzmann equation for the non-condensate. Despite its successes, however, this approach neglects the anomalous average and all Beliaev processes. Although Beliaev processes are expected to be swamped by Landau processes at high temperature, they have nonetheless been directly observed in a number of recent experiments \cite{Hodby01,Katz02,Bretin03,Mizushima03}, while the good agreement with the JILA results for the $m=2$ mode within the GHFB theory \cite{Hutchinson98} suggests that the anomalous average can also be important.

In two recent papers, we have developed a systematic perturbative extension of the Bogoliubov theory which includes these effects and explains the JILA results for both the $m=2$ and $m=0$ modes \cite{Morgan03a,Morgan04}. The formalism adapts the linear response treatment of Giorgini \cite{Giorgini00} and provides a time-dependent extension of an earlier second-order perturbative calculation \cite{Morgan00}. The theory is gapless and includes the dynamic coupling between the condensate and non-condensate, all relevant Beliaev and Landau processes and the anomalous average. It is also consistent with the generalized Kohn theorem for the dipole modes \cite{Dobson94}. The theory is valid in the collisionless limit of well-defined quasiparticles, which requires $(k_{\text{B}}T/n_{0}U_{0})(n_0a_s^{3})^{1/2} \ll 1 $ where $n_0$ is the condensate density, $a_s$ is the s-wave scattering length, $k_{\text{B}}$ is Boltzmann's constant and $U_0 = 4\pi\hbar^2a_s/m$ where $m$ is the atomic mass \cite{Giorgini00,Morgan00}. For the JILA experiment \cite{Jin97} this parameter does not exceed $0.03$ at the trap centre for the highest temperature we consider.

An important feature of our theoretical approach is that it explicitly includes the effect of the external perturbation on the non-condensate dynamics. At high temperatures and for perturbations which are peaked on the edge of the atomic cloud, the non-condensate has a large response arising from single-particle resonances at integer values of the trap frequencies. If the quasiparticle mode is located close to such a resonance, it is possible to excite the condensate via the thermal cloud as intermediary and the strength of this excitation can exceed the direct effect of the external perturbation. This turns out to be the explanation for the anomalous behaviour of the $m=0$ mode in the JILA experiment \cite{Morgan03a} and it is also necessary to include this process for a correct description of the dipole oscillations of the system, as we show later in this paper.

\subsection{Outline of the paper}

This paper is organised as follows. In Sec.~\ref{sec:theory:summary} we summarize the relevant results of the theory developed in Ref.~\cite{Morgan04} and in Sec.~\ref{sec:numerical:method} and its subsections we discuss our numerical methods. We focus particularly on the important issue of convergence and in Sec.~\ref{sec:convergence} we describe a new asymmetric summation technique we have recently developed which greatly improves the convergence properties of the calculation. In Sec.~\ref{sec:numerical:results} we present our numerical results for the parameters of the JILA experiment. We focus first on the results for the $m=2$ and $m=0$ modes, describing the effect of some of the Landau and Beliaev processes which can take place and showing the importance of including the effect of the perturbation on the thermal cloud. In Sec.~\ref{sec:results:N0} we analyse the experimental results for the condensate population and temperature to determine the correct input parameters for the theory and in Sec.~\ref{sec:dipole} we provide results for the dipole modes and show that these are obtained to high accuracy in our calculation. Finally, in Sec.~\ref{sec:phase} we present results for the relative phase of the condensate--non-condensate oscillations and demonstrate that the cross-over from out-of-phase to in-phase oscillations predicted for the $m=0$ mode at high temperature by Al Khawaja, Bijlsma and Stoof \cite{Bijlsma99,Khawaja00} is reproduced in our theory. Two appendices contain details of the calculation and the definitions of some of the quantities we require.

\section{Summary of second order Beliaev-Popov theory} \label{sec:theory:summary}

In this section we briefly summarize the theory that we use in this paper, which we will refer to as the second order Beliaev-Popov (SOBP) theory. The full derivation of the expressions we quote along with a detailed discussion of their physical significance is given in Ref.~\cite{Morgan04}, to which we frequently refer.

\subsection{Equations of motion and Bogoliubov quasiparticles}

The SOBP theory describes the evolution of a condensate in the presence of non-condensed atoms using a generalized GPE containing various non-condensate mean-fields. These are calculated by modelling the thermal cloud as a non-interacting gas of quasiparticles evolving in the time-dependent mean-field of the condensate according to an appropriate set of Bogoliubov-de Gennes (BdG) equations. Together the GPE and BdG equations form a consistent description of the coupled dynamics of the system which contains all the leading order corrections to the usual Bogoliubov theory in the relevant small parameter, defined above.

The generalized GPE for the condensate wavefunction $\Phi\rt$ (normalized to one) has the form
\begin{align}
 i \hbar \frac{\partial}{\partial t} \Phi\rt &= \left[ \hsp(\bfr) + P\rt - \lambda(t)\right ] \Phi\rt \nonumber \\
&+ \left[ N_0(t)U_0|\Phi\rt|^2 + 2 U_0 \nt\rt \right ] \Phi\rt \nonumber \\
&+ U_0 \mt^{\text{R}}\rt \Phi^*\rt - f\rt, \label{TDGGPE}
\end{align}
where $\hsp(\bfr) = -\hbar^2 \nabla^2/2m + V_{\text{trap}}(\bfr)$ is the single-particle Hamiltonian containing the static trap potential (if any), $P\rt$ is a time-dependent external perturbation, $\lambda(t)$ is a scalar which controls the global phase of $\Phi\rt$, and $N_0(t)$ is the condensate population \cite{dropDN0}. Particle interactions are described using a contact potential with interaction strength $U_0$. The use of this contact potential necessitates an ultra-violet (UV) renormalization of the theory, which is achieved by a suitable modification of the pair average $\mt^{\text{R}}\rt$ (see below).

The coupling of condensed and non-condensed atoms is described by the terms involving the non-condensate density $\nt\rt$, the renormalized anomalous (pair) average $\mt^{\text{R}}\rt$, and the function $f\rt$ which arises from the fact that the wavefunctions of the non-condensed atoms are orthogonal to the condensate. These quantities are constructed from a complete set of time-dependent quasiparticle wavefunctions $u_i\rt$ and $v_i\rt$ according to
\begin{align}
\nt \rt &= \sum_{i} |u_i\rt|^2 N_i +  |v_i\rt|^2 (N_i + 1), \label{nt_qp}\\
\mt^{\text{R}} \rt &= \sum_{i} u_i\rt v_i^*\rt (2N_i + 1)+ \frac{N_0\Delta U_0}{U_0} \Phi^2\rt,\label{mt_qp} \\
f\rt &= \frac{1}{N_0}\sum_{i} c_i^*N_i u_i\rt + c_i(N_i+1)v_i^*\rt,\\
c_i(t) &= N_0U_0 \!\! \intdr |\Phi|^2 \left [ \Phi^* u_i\rt + \Phi v_i\rt \right ].\label{ci}
\end{align}
The final term in the expression for $\mt^{\text{R}}\rt$ is the UV-renormalization and the quantity $\Delta U_0$ is the second-order approximation to the interaction strength $U_0$ as calculated from the Lippmann-Schwinger equation
\beq
\Delta U_0 = U_0^2 \int \frac{d^3{\bf k}}{(2\pi)^3} \frac{1}{2(\hbar^2k^2/2m)}.
\label{DeltaU0}
\eeq

To obtain a closed set of equations we also require the equation of motion for the quasiparticle wavefunctions. These evolve according to the BdG equations
\begin{gather}
i \hbar \frac{\partial}{\partial t} \bpm u_{i} \\ v_{i} \epm = \bpm \hat{L} & \hat{M} \\ -\hat{M}^* & -\hat{L}^* \epm\bpm u_{i} \\ v_{i} \epm,\label{uvt}\\
\hat{L}\rt = \hsp + P\rt +N_0U_0 \left [ |\Phi|^2 + \hat{Q}|\Phi|^2\hat{Q} \right ],\label{Lt}\\
\hat{M}\rt = N_0 U_0\hat{Q}\Phi^2\hat{Q}^*,\label{Mt}
\end{gather}
where the projector $\hat{Q} = 1-|\Phi\rangle\langle \Phi|$ ensures orthogonality of $\Phi\rt$ and $\{u_i\rt,v_i^*\rt\}$. The quasiparticle populations $\{ N_i \}$ are independent of time and given by the Bose-Einstein distribution $N_i = 1/(e^{\beta(\epsilon_i-\delta\mu)}-1)$, where $\epsilon_i$ is the Bogoliubov energy (see below), $\beta = 1/k_{\text{B}}T$ and $\delta\mu$ is the (small) difference between the condensate energy and the chemical potential. The condensate population is defined implicitly using the constraint on the total particle number $N$ by $N_0(t) = N-\intdr \nt\rt$. Most quantities in the theory depend on temperature via their dependence on the quasiparticle populations. 

The above equations are obtained using the number-conserving formalism of Gardiner and Castin and Dum, modified for finite temperature calculations \cite{Gardiner97,Castin98,Morgan04,Morgan03d}. The terms $f\rt$ and $\hat{Q}$ which arise from orthogonality of the condensate and non-condensate are a feature of this approach and do not appear in symmetry-breaking theories. We find numerically that they can give a significant contribution to the energy shifts (see Sec.~\ref{sec:finitesize}).

\subsection{Equilibrium solutions}

In equilibrium, $P\rt = 0$ and Eq.~(\ref{TDGGPE}) has a time-independent solution $\Phi\rt = \Phi(\bfr)$ which satisfies
\begin{align}
 \left[ \hsp(\bfr) - \lambda + N_0U_0|\Phi(\bfr)|^2 + 2U_0\nt(\bfr)\right] \Phi(\bfr)& \label{TIGGPE}\\
+ U_0\mt^{\text{R}}(\bfr)\Phi^*(\bfr) - f(\bfr)& = 0, \nonumber
\end{align}
where $\lambda$ is the condensate eigenvalue and $\nt(\bfr)$, $\mt^{\text{R}}(\bfr)$ and $f(\bfr)$ are equilibrium non-condensate mean-fields calculated from Eqs.~(\ref{nt_qp})-(\ref{ci}) using the static quasiparticle basis defined below. Setting these quantities to zero gives the usual time-independent GPE with wavefunction $\cond$ and energy $\lambda_0$
\beq
\left[ \hsp(\bfr) - \lambda + N_0U_0|\cond|^2 \right] \cond = 0.
\label{TIGPE}
\eeq
We solve Eq.~(\ref{TIGGPE}) by linearizing the change in energy and shape of the condensate relative to this solution.

We obtain time-independent BdG equations by writing $\Phi\rt \rightarrow \cond$ and $P\rt = 0$ in Eqs.~(\ref{uvt})-(\ref{Mt}) and looking for solutions of the form
\beq
\bpm u_i\rt \\ v_i\rt \epm = \bpm u_i(\bfr) \\v_i(\bfr) \epm e^{-i\epsilon_i t/\hbar}.
\label{qpBasis}
\eeq
The equilibrium Bogoliubov quasiparticle wavefunctions and energies are therefore found by solving
\beq
\bpm \hat{L}_0 & \hat{M}_0 \\ -\hat{M}_0^*  &-\hat{L}_0^* \epm\bpm u_{i} \\ v_{i} \epm = \epsilon_i \bpm u_{i} \\ v_{i} \epm,
\label{BdG}
\eeq
where $\hat{L}_0$ and $\hat{M}_0$ are defined by making the appropriate substitutions in Eqs.~(\ref{Lt})-(\ref{Mt}). An important consequence of the projector $\hat{Q}$ in $\hat{L}_0$ and $\hat{M}_0$ is the existence of two zero-energy solutions proportional to $\cond$ which means that these quasiparticle wavefunctions form a complete basis set. In particular, we obtain a set of solutions
\beq
\left \{ \bpm u_i \\ v_i \epm, \bpm v_i^* \\ u_i^* \epm, \bpm \Phi_0 \\ 0 \epm, \bpm 0 \\ \Phi_0^* \epm\right \},
\eeq 
with energies $\epsilon_i, -\epsilon_i, 0, 0$ and norms $+1,-1,+1,-1$ respectively, where the norms are defined by $\intdr |u_i|^2 - |v_i|^2$. We use the notation $i = 0+$ to denote the positive-norm, zero-energy mode, which we will also refer to as `the zero mode'. The ease with which the two zero energy solutions are obtained and used in the theory is an important advantage of the number-conserving approach.

\subsection{Linear response theory}

We can use the above equations to find the linear response of a Bose-condensed gas to an external perturbation. We consider the situation where a condensate has been formed at low temperature and has settled into the ground state of Eq.~(\ref{TIGGPE}) \cite{cite:groundstate}. When the external perturbation $P\rt$ is applied, the system responds with a time-dependent oscillation of all mean fields around their static values, $\Phi \rt = \Phi(\bfr) + \delta \Phi\rt$, $\nt \rt = \nt(\bfr) + \delta\nt\rt$ etc. Substituting these expressions into Eq.~(\ref{TDGGPE}) and linearizing leads to the equation of motion for the condensate fluctuation $\delta \Phi \rt$. This equation can be solved by combining it with its complex conjugate, Fourier transforming and expanding in the equilibrium quasiparticle basis 
\beq
\coldgPw = \sum_{i} b_i(\omega) \col{u_i}{v_i}.
\label{qp_expansion}
\eeq
The expansion coefficients $b_i(\omega)$ (response amplitudes) are directly related to the condensate density fluctuations $\delta n_c = \delta (N_0|\Phi|^2)$, which are the experimentally relevant quantities
\beq
\delta n_c({\bf r},\omega) = \delta N_0(\omega)|\Phi_0|^2 + N_0 \!\!\sum_{i} \! b_i(\omega)\bigl [\Phi_0^* u_i + \Phi_0 v_i \bigr ],
\label{dnc_GGPE}
\eeq
where $\delta N_0(\omega) = -\intdr \delta \nt\rw$ describes any fluctuations in the condensate population. This quantity is only non-zero if the perturbation has a contribution with the same symmetry as the condensate (as for the $m=0$ mode in the JILA experiment for example) and we have found numerically that it generally has rather little effect on the observed density fluctuations.

In general, the expansion coefficients $b_i(\omega)$ can be found from the solution of a linear matrix problem. The result is particularly simple, however, if a single positive-norm mode dominates the expansion, which is usually the situation for experiments designed to study excitations. In the case that mode `p' is excited, the solution has the form
\beq
b_p(\omega) = P_{p0}(\omega)\mathcal{F}_p(\omega+i\gamma),
\label{bpF}
\eeq
where $\gamma$ measures the experimental resolution, the excitation matrix element $P_{p0}(\omega)$ is defined by
\beq
P_{p0}(\omega) = \intdr P \rw \left [ u_{p}^{*} \Phi_0 + v_{p}^{*} \Phi_0^* \right ],\label{P_p0}
\eeq
and the response function (resolvent) $\mathcal{F}_p(\omega)$ takes different forms depending on the level of approximation. In the simplest case where the coupling to the non-condensate is completely neglected, $\mathcal{F}_p(\omega)$ is a  lorentzian centred on the frequency of the corresponding Bogoliubov mode
\beq
\mathcal{F}_p(\omega) = \frac{1}{\hbar \omega-\epsilon_p}.
\label{F}
\eeq
This function diverges at the resonance frequency $\omega_p = \epsilon_p/\hbar$ because there is no intrinsic damping in the theory at this level of approximation. The inclusion of the resolution parameter $\gamma$ in Eq.~(\ref{bpF}) via the standard substitution $\omega \rightarrow \omega + i \gamma$ ensures we obtain finite quantities (essential for numerical work) and can be justified from the finite experimental observation time \cite{Morgan04}. Typically, $\gamma$ is of order a few hundredths of an appropriate trapping frequency and our results do not depend strongly on its precise value within the experimentally relevant range.

If we now include the dynamic coupling between the condensate and non-condensate, the response function becomes 
\begin{gather}
\mathcal{F}_p(\omega) = \mathcal{R}_p(\omega),
\label{FR}
\\
\mathcal{R}_p(\omega) = \mathcal{G}_p(\omega) + \tilde{\mathcal{G}}_p(\omega),\label{R}\\
\tilde{\mathcal{G}}_p(\omega) = \left [ \frac{\Delta P_{p0}^{(S)}(\omega) + \Delta P_{p0}^{(D)}(\omega)}{P_{p0}(\omega)} \right ]\mathcal{G}_p(\omega),
\label{Gt}
\\
\mathcal{G}_p(\omega) = \frac{1}{\hbar \omega - E_p\omega)},\label{G}\\
E_p(\omega) = \epsilon_p + \Sigma_p(\omega). \label{Epw}
\end{gather}
Here $\Sigma_p(\omega)$ is a frequency-dependent self-energy, while the quantities $\Delta P_{p0}^{(S)}(\omega)$ and $\Delta P_{p0}^{(D)}(\omega)$ depend on the functional form of the external perturbation and describe changes in the excitation amplitude $P_{p0}(\omega)$ resulting from the coupling of the condensate to the thermal cloud. 

The various response functions introduced above describe different dynamical processes occurring in the gas. In general both the condensed and uncondensed atoms can be excited via two distinct mechanisms; either directly by the external perturbation or indirectly by fluctuations in the other mean-fields. For the case of the condensate, these two mechanisms are described by the response functions $\mathcal{G}_p(\omega)$ and $\tilde{\mathcal{G}}_p(\omega)$ respectively, while the total response including both processes is described by $\mathcal{R}_p(\omega)$. 

The separate response functions $\mathcal{G}_p(\omega)$ and $\tilde{\mathcal{G}}_p(\omega)$ are introduced partly to facilitate interpretation of the theory and partly because they may dominate in certain situations. In general, the relatively large population and density of the condensate means that the dominant process at low temperature is for the perturbation to excite the condensate which subsequently drives the non-condensed atoms via their mutual coupling. This case is described by $\mathcal{G}_p(\omega)$ alone. The alternative mechanism where the perturbation excites the non-condensate first and this acts on the condensate in a second step is described by $\tilde{\mathcal{G}}_p(\omega)$. Under certain circumstances, (and especially at finite temperatures) this second process can become dominant and its inclusion is crucial to explain the JILA experiment and for an accurate description of the dipole modes, as we show in Sec.~\ref{sec:dipole} \cite{Jin97,Morgan03a}. It is also possible to enhance the effect of one or other of these mechanisms by a suitable choice of the perturbing potential $P\rt$. If this is localized on the condensate then the response will be dominated by $\mathcal{G}_p(\omega)$, while if it mainly acts in the wings of the non-condensate then $\tilde{\mathcal{G}}_p(\omega)$ is more appropriate. This later case has been used at MIT to investigate second sound oscillations \cite{StamperKurn98}. We stress, however, that the full response is given by $\mathcal{R}_p(\omega)$ and includes both mechanisms.

The self-energy in Eq.~(\ref{Epw}) contains two distinct types of contributions, static $(S)$ and dynamic $(D)$, corresponding to the different roles of the thermal cloud
\beq
\Sigma_p(\omega) = \Delta E_{pp}^{(S)} + \Delta E_{pp}^{(D)}(\omega).
\label{sigma}
\eeq
The frequency-independent static term $\Delta E_{pp}^{(S)}$ comes from interactions between a condensate fluctuation and the equilibrium non-condensate mean-fields, while the dynamic term $\Delta E_{pp}^{(D)}(\omega)$ describes the driving of the non-condensed atoms by the condensate and their subsequent back action.

The detailed definition of these quantities is given in \cite{Morgan04} and in Appendix~\ref{sec:app:shifts}. Briefly, the static term can be written as a sum of contributions of the form
\beq
\Delta E_{pp}^{(S)} = \Delta E_{4}(p) + \Delta E_{\lambda}(p) + \Delta E_{\text{shape}}(p) + \Delta E_{f}^{(S)}(p). \label{DEs}
\eeq
Here $\Delta E_{4}(p)$ and $\Delta E_{f}^{(S)}(p)$ arise from the explicit interactions between a condensate fluctuation and the equilibrium mean-fields $\nt(\bfr)$, $\mt^{\text{R}}(\bfr)$ and $f(\bfr)$, while $\Delta E_{\lambda}(p)$ and $\Delta E_{\text{shape}}(p)$ come from the effect these mean-fields have on the energy and shape of the equilibrium condensate [i.e. the difference between the solutions to Eqs.~(\ref{TIGGPE}) and (\ref{TIGPE})]. All static terms can be calculated straightforwardly from integrals involving the condensate wavefunction and the equilibrium non-condensate mean-fields. The dynamic terms on the other hand involve a double summation over the quasiparticle basis states of the form
\beq
\Delta E_{pp}^{(D)}, \Delta P_{p0}^{(D)} \sim \sum_{ij} \frac{f(N_i)M_{pij}}{\omega -\omega_{ij} + i\gamma}.
\label{dD}
\eeq
Here $f(N_i)$ is a simple function of the quasiparticle populations $N_i$, $M_{pij}$ is a suitable matrix element and $\omega_{ij}$ is a resonance of the non-condensate corresponding to a Beliaev or Landau process ($\omega_{ij} = \pm(\epsilon_i+\epsilon_j)$ and $\omega_{ij} = \pm(\epsilon_i - \epsilon_j)$ respectively). The matrix elements $M_{pij}$ are the product of two factors each of which is defined in terms of integrals of the equilibrium condensate wavefunction $\cond$ and the quasiparticle wavefunctions for modes $p$, $i$, and $j$.

\section{Numerical method} \label{sec:numerical:method}

We have used the formalism outlined above to calculate the density response of a trapped condensate as a function of temperature. In this section we describe the numerical techniques required and present some illustrative results. The calculation is difficult because some of the intermediate terms we require are much larger than the final self-energy so there is substantial cancellation between them. This puts significant demands on numerical accuracy and convergence which we have solved using a variety of techniques, described briefly below. Further details of the methods used can be found in Ref.~\cite{Morgan05b}.

We consider the case that the trapping potential is an axisymmetric, anisotropic harmonic potential of the form
\beq
V_{\text{trap}}(\bfr) = \frac{1}{2}m(\omega_{r}^2r^2 + \omega_{z}^2z^2),
\eeq
where $r^2 = x^2+y^2$ is the square of the radial coordinate. In the present paper we use simulation parameters appropriate to the 1997 JILA experiment \cite{Jin97}. The trap frequencies are therefore $\nu_{r} = \omega_{r}/(2\pi) = 129$Hz and $\nu_{z} = \omega_{z}/(2\pi) = 365$Hz, the scattering length for the Rb$^{87}$ atoms is $a_s = 110a_0$ where $a_0$ is the Bohr radius and the resolution parameter $\gamma = 0.36\omega_r$ \cite{Jin96,Jin97}. These parameters are fixed for all the results presented in this paper. In addition we take the condensate population $N_0$ to be $6000$ unless specifically stated otherwise.

For a given atomic species and trap geometry, the variable input parameters of the theory are the condensate population $N_0$ and the absolute temperature $T$. For a given value of $N_0$, the numerical calculation of the condensate density response can be broken down into the following steps:
\begin{enumerate}
\item Solve the time-independent GPE of Eq.~(\ref{TIGPE}) to obtain the static condensate wavefunction $\cond$ and eigenvalue $\lambda_0$.

\item Solve the equilibrium BdG equations of Eq.~(\ref{BdG}) to obtain the static quasiparticle wavefunctions $u_i(\bfr)$ and $v_i(\bfr)$ and energies $\epsilon_i$ for all states up to a numerical cutoff energy $E_{cut}$. These solutions are saved to disk along with the condensate wavefunction and eigenvalue. 

\item For each temperature $T$, construct the equilibrium non-condensate density $\nt(\bfr)$ and renormalized anomalous average $\mt^{\text{R}}(\bfr)$ on a spatial grid by a direct summation over the quasiparticle states using the time-independent form of Eqs.~(\ref{nt_qp}) and (\ref{mt_qp}), supplemented by a semi-classical approximation for states above the numerical cutoff (see Appendix~\ref{sec:app:semi-classical}). From these, the equilibrium solutions to the generalized GPE can be found straightforwardly, as can all the static shifts defined in Eq.~(\ref{DEs}). These calculations simply involve integrations over the spatial grids defining the equilibrium mean-fields.

\item For each mode `p', temperature $T$ and resolution $\gamma$ the dynamic terms $\Delta E_p^{(D)}(\omega)$ and $\Delta P_{p0}^{(D)}(\omega)$ (and $\delta N_0(\omega)$ if required) are calculated on a frequency grid centred on the Bogoliubov energy $\epsilon_p$. This requires looping over the double quasiparticle summation, calculating the non-zero matrix elements for the relevant Landau and Beliaev processes and combining these with the appropriate energy denominator. The magnitude of the task is greatly reduced by selection rules which mean that only a relatively small fraction of the terms in the sum are non-zero. For $\Delta P_{p0}^{(D)}(\omega)$, this step requires a choice for the functional form of the external perturbation $P\rt$, and this is taken to mimic the experiment as closely as possible [see Eq.~(\ref{perturbation})]. In some cases a semi-classical approximation is also used to include the effect of high energy states above the numerical cutoff.

\item The various contributions are combined to give the self-energy $\Sigma_p(\omega+i\gamma)$ and the response functions $\mathcal{G}_p(\omega+i\gamma)$, $\tilde{\mathcal{G}}_p(\omega+i\gamma)$ and $\mathcal{R}_p(\omega+i\gamma)$. Energies and decay rates are extracted from these quantities either by reading off the value of the self-energy at the poles or by fitting appropriate functions to the resolvents in either the time or frequency domains.

\item To compare with the experimental data, our results must be plotted against the reduced temperature $t = T/T_c^0$ rather than the absolute temperature $T$. Here $T_c^0$ is the critical temperature of an ideal gas of $N$ atoms in a harmonic trap, given by
\beq
\frac{k_{\text{B}}T_c^0}{\hbar \bar{\omega}} = \left(\frac{N}{\zeta(3)}\right)^{1/3},
\label{Tc0}
\eeq
where $\bar{\omega}=(\omega_{r}^2\omega_z)^{1/3}$ is the geometric mean trap frequency, $\zeta(x)$ is the Riemann zeta function, and $\zeta(3) \approx 1.202$. The conversion requires calculating the total atom number $N$ for each temperature, which is achieved using the relation $N = N_0 + \intdr \nt(\bfr)$. This gives $N$, $T_c^0$ and $t$ as a function of temperature in nanokelvin for a given fixed value of the condensate population $N_0$. 
\end{enumerate}

Typically, we solve the BdG equations with a very large energy cutoff of $E_{cut} \sim 140 \hbar \omega_r$ which corresponds to more than $100,000$ quasiparticle modes. This takes about $5$ hours of computation on a desktop PC with a 2.4GHz Pentium 4 CPU and 1Gb RAM and consumes just over $2$Gb of storage on disk. The subsequent calculation of all the relevant response functions for $20$ different temperatures and a frequency grid of 512 points takes roughly an additional $5$ hours of computation for each excitation under study. In the present work these are restricted to the lowest energy $m=2$ and $m=0$ modes and the dipole oscillations.

In the following sections, we provide further details on some of the above steps along with illustrative numerical results.

\subsection{GPE solution}

The GPE of Eq.~(\ref{TIGPE}) is solved by expanding the wave function $\cond$ in an appropriate basis and solving the resulting set of coupled nonlinear equations for the basis coefficients. The basis states are chosen to be products of the harmonic oscillator (HO) states for the radial and axial directions, while the solution of the nonlinear equations is achieved using the MATLAB optimisation routine \texttt{FSOLVE}, with an initial guess provided by the Thomas-Fermi solution. This is an efficient method of solving the GPE as it is relatively stable initially while also giving rapid (quadratic) convergence near the end of the solution cycle. Various integrals of products of four functions (such as the projection of $|\Phi_0|^2\Phi_0$ onto a basis state for example) are required to set up the nonlinear problem. These are calculated on a Gaussian quadrature grid chosen to allow exact integration of such products.

\subsection{BdG solution}

The equilibrium BdG equations given in Eq.~(\ref{BdG}) are solved using the procedure described in Ref.~\cite{Hutchinson00}. The symmetry of the trap potential means that the equations decouple into subspaces with definite values for the $z$-component of angular momentum $m$ and \textit{z}-parity $p = 0$ (even) or $1$ (odd). Within each subspace, the solutions are assigned another quantum number $n$ which orders the energy within that subspace. The subscript `i' on the quasiparticle wavefunctions $u_i(\bfr)$ and $v_i(\bfr)$ therefore stands for the triplet of quantum numbers $(m_i,p_i,n_i)$. The quasiparticle energies and the radial and axial parts of the wavefunctions only depend on the modulus of $m_i$ so we only need to solve the BdG equations in subspaces with $m_i \geq 0$. The dependence on the angular coordinate $\phi$ has the usual complex exponential form and the solutions can therefore be written as
\begin{align}
u_i(\bfr) &= u_{|m_i|,p_i,n_i}(r,z)\frac{e^{im_i\phi}}{\sqrt{2\pi}},\\ 
v_i(\bfr) &= v_{|m_i|,p_i,n_i}(r,z)\frac{e^{im_i\phi}}{\sqrt{2\pi}}.
\end{align}

Within a given subspace of $m$ and \textit{z}-parity, the BdG equations are solved in two stages. In the first stage, we obtain a single-particle basis orthogonal to the condensate and with the appropriate symmetries by solving the equation
\beq
 \left [ \hsp(\bfr) + N_0U_0 | \Phi_0 \rt|^2 \right ] \zeta_i(\bfr) = \epsilon_i^{\scriptscriptstyle (GP)} \zeta_i(\bfr).
 \label{GPE_basis}
\eeq
We will refer to the solutions of this equation with $\epsilon_i^{\scriptscriptstyle (GP)}\neq \lambda_0$ as the GPE basis. 
Equation~(\ref{GPE_basis}) is solved by expanding the wave functions $\{\zeta_i\}$ using the underlying HO basis set. The BdG equations are solved by rewriting them as decoupled equations for $u_i \pm v_i$ and expanding in the GPE basis which reduces them to a standard matrix eigenvalue problem. Finally, the solutions are converted from the GPE basis to the HO basis and stored. The GPE basis wave functions are introduced for two reasons; firstly, they are orthogonal to the condensate and therefore the treatment of the orthogonal projector is trivial and secondly the BdG equations take a particularly simple form when written in the GPE basis \cite{Hutchinson00}. 

We solve for all quasiparticle modes with energies less than a numerical cutoff $E_{cut}$ determined by setting a maximum value for the angular momentum quantum number $m$, which we denote by $m_{max}$. $E_{cut}$ is then taken as the lowest energy state in the subspace with $m = m_{max}$ and even \textit{z}-parity. In our largest calculations we take $m_{max} = 150$ which corresponds to $E_{cut} \approx 140 \hbar \omega_{r}$. Such a large basis set is necessary to ensure numerical convergence of our final results (see Sec.~\ref{sec:convergence}) and requires that care be taken to maintain numerical accuracy at all stages of the calculation.

The solution of the BdG equations requires the calculation of numerous integrals involving products of either two or four functions. Integrals of two functions can be done exactly using the scalar product of the HO expansion coefficients. Integrals of four functions are calculated on a Gaussian quadrature grid, as in the solution of the GPE. The grid points and weights are chosen to allow exact integration of any product of four functions constructed from the underlying HO basis. However, since all the integrals required contain the condensate density, we discard all points which lie outside the region where this has fallen to $10^{-10}$ of its peak value. This has essentially no effect on the accuracy of the integration but keeps the numerical grid to a reasonable size for calculations with a large basis.

\subsection{Matrix elements for dynamic terms}

The dynamic terms require the calculation of numerous matrix elements and these are produced `on-the-fly' by loading the stored quasiparticle wavefunctions in the HO representation, constructing them on a suitable spatial grid and summing over the points. The number of integrals which must be evaluated is kept manageable using angular momentum and \textit{z}-parity selection rules which mean that only a small subset of the matrix elements are non-zero. For the dynamic self-energy $\Delta E_p^{(D)}(\omega)$, the integrals involve products of four wavefunctions with two of low energy (the condensate wave function and the quasiparticle mode `p' under study) and two with energies up to the numerical cutoff $E_{cut}$. These integrals can therefore be done essentially exactly using the truncated Gaussian quadrature grid used in the solution of the BdG equations.

Unfortunately, this grid can not be used to calculate the matrix elements required for $\Delta P_{p0}^{(D)}(\omega)$ which involve the product of the perturbation $P\rw$ in the space-frequency domain and two quasiparticle wavefunctions. The reason is that $P\rw$ is generally not localized in the centre of the condensate but extends out to the wings of the trapped cloud. However, the perturbation typically involves only very low order polynomials so these integrals can instead be evaluated by constructing a new Gaussian quadrature grid which can integrate bilinear products of quasiparticle functions exactly and then adding the few extra points required to deal with the perturbation. In this way all integrals required in the theory are calculated exactly and the accuracy of the numerics is limited purely by the size of the basis employed (i.e. by the value of $E_{cut}$ and, for a given $E_{cut}$, by the size of the HO basis used to construct the solutions).

The functional form of the perturbation is chosen to mimic the experiment as closely as possible. In general, we take
\beq
P\rt = \frac{1}{\pi}P_p(r,z)\cos(\omega_d t-m_p\phi)\Theta(t)\Theta(t_d-t),
\label{perturbation}
\eeq
where $\omega_d$ is the central drive frequency, $\Theta(t)$ is the unit step function, and the perturbation is applied for $0 < t < t_d$ ($t_d \approx 14$ms in the JILA experiment \cite{Jin97}). Following the form of the perturbation used experimentally, we take $P_p(r,z) \propto r^2$ for the $m=0$ and $m=2$ modes, while $P_p(r,z) \propto z$ for the dipole mode along the \textit{z}-axis and $P_p(r,z) \propto r$ for the dipole modes in the $x$-$y$ plane.

\subsection{Dynamic Summations}

The dynamic terms are calculated by performing the double sum over intermediate quasiparticle states $i$ and $j$ in Eq.~(\ref{dD}). The calculation is only feasible because the selection rules on the matrix elements mean that only a small fraction of the total number of intermediate pairs contribute in the sum. In fact, for a given quasiparticle mode `p', each value of the angular momentum $m_i$ and \textit{z}-parity $p_i$ for mode $i$ corresponds to unique values for $m_j$ and $p_j$, so the calculation can be broken into subspace blocks. For each block, the quasiparticle energies and wavefunctions are loaded and the relevant matrix elements are calculated as described above. The summations over the remaining quantum numbers $n_i$ and $n_j$ (describing the energies within each block) are then performed for each temperature $T$ and resolution parameter $\gamma$ and for each value of $\omega$ on a frequency grid centred on the energy of the mode under study. This process is repeated for all contributing subspace blocks and for all modes `p' of interest.

The efficiency of this calculation can be greatly increased by noting that the frequency dependence of the dynamic terms generally consists of a few strong features superposed on a smooth background. A pair of intermediate states $i$ and $j$ is associated with an energy resonance for a Beliaev and Landau process when $\hbar\omega = \pm \epsilon_i \pm \epsilon_j$ [cf. Eq.~(\ref{dD})]. If this resonance occurs within the frequency range of experimental interest, then it potentially corresponds to a sharp feature which in principle we need to resolve. This situation therefore requires a fine grid spacing, although generally the sum of many such terms is much smoother than each individually. The fine grid is needed, however, if a few modes dominate the response, a situation which does occur in finite systems (as our results demonstrate) and which has been termed `temperature induced resonances' by Guilleumas and Pitaevskii \cite{Guilleumas99}. However, the vast majority of resonances fall outside the frequency range of direct interest (for example there is usually at most one Beliaev resonance within this range for typical values). The contribution from these processes is smooth in frequency and can be adequately represented on a rather coarse grid. We therefore divide the intermediate states into two groups; those with resonances in the range of interest are calculated on a fine grid while those with resonances outside this range are calculated on a much coarser grid and interpolated onto the fine grid at the end of the calculation. This scheme dramatically reduces the computational time at a negligible cost in accuracy. In the calculations presented here we used a fine grid of $512$ points over a frequency range of $2\omega_r$ with a coarse grid of only $16$ points over the same range.

\subsection{Convergence} \label{sec:convergence}

An important requirement of any numerical calculation is that the results have converged sufficiently that meaningful conclusions can be drawn. This is particularly difficult in the present case because of residual effects of the infrared divergence problem which plagues theories of the Bose gas beyond the Bogoliubov approximation. Although the full theory is infrared finite, individual terms in the self-energies are not; for example, in the large volume homogeneous limit both the static and dynamic contributions diverge as $1/k$ as $k \rightarrow 0$ and only their sum is finite and proportional to the small parameter of the theory \cite{Fedichev98,Giorgini00,Morgan00}.

In a finite system, this divergence is suppressed but the static and dynamic terms may still become large compared to other energy scales in the problem. This is demonstrated in Table~\ref{table:energies} which shows the values of various contributions to the self-energy for a range of reduced temperatures.
As in the homogeneous case, the removal of the infrared divergence can be seen in the substantial cancellation between the total static shift $\Delta E^{(S)}$ and the dynamic contribution $\Delta E^{(D)}(\omega)$, while the small remaining difference $\Delta E$ (of order a tenth of a trap frequency or less) represents the overall change in energy compared to the Bogoliubov theory.

\begin{table*}
\begin{ruledtabular}
\begin{tabular}{ddddddddd}
\multicolumn{1}{c}{$t$} & \multicolumn{1}{c}{$\Delta E_{4}$} & \multicolumn{1}{c}{$\Delta E_{\lambda}$} & \multicolumn{1}{c}{$\Delta E_{\text{shape}}$} & \multicolumn{1}{c}{$\Delta E_{f}^{(S)}$} & \multicolumn{1}{c}{$\Delta E^{(S)}$} & \multicolumn{1}{c}{$\Delta E^{(D)}(\omega = \epsilon_p/\hbar)$} & \multicolumn{1}{c}{$\Delta E_{0}$} & \multicolumn{1}{c}{$\Delta E$} \\ \hline
0.00 & -0.01 &  -0.19  &  0.05  &  0.00 &  -0.14 &  0.16 & 0.00 &  0.01 \\
0.10 &  0.11 &  -0.24  &  0.04  &  0.00 &  -0.09 &  0.11 & 0.00 &  0.02 \\
0.20 &  0.38 &  -0.38  &  0.02  & -0.01 &   0.00 &  0.01 & 0.01 &  0.01 \\
0.30 &  0.75 &  -0.61  & -0.01  & -0.01 &   0.12 & -0.11 & 0.01 &  0.01 \\
0.40 &  1.26 &  -0.95  & -0.04  & -0.02 &   0.26 & -0.26 & 0.01 &  0.00 \\
0.50 &  1.93 &  -1.41  & -0.07  & -0.02 &   0.42 & -0.43 & 0.02 & -0.01 \\
0.60 &  2.88 &  -2.09  & -0.12  & -0.03 &   0.63 & -0.65 & 0.02 & -0.02 \\
0.70 &  4.35 &  -3.20  & -0.19  & -0.04 &   0.92 & -0.96 & 0.03 & -0.04 \\
0.80 &  7.20 &  -5.42  & -0.30  & -0.06 &   1.42 & -1.49 & 0.04 & -0.07 \\
0.90 & 16.38 & -12.94  & -0.60  & -0.10 &   2.75 & -2.89 & 0.06 & -0.14
\end{tabular}
\end{ruledtabular}
\caption{Contributions to the self-energy for the $m = 2$ mode and a range of reduced temperatures $t$ for the parameters of the JILA TOP trap and a condensate population of $N_0 = 6000$. All terms include a semi-classical contribution from high energy states, UV renormalization (if appropriate) and are given to two decimal places in units of $\hbar \omega_r$. The various contributions to the static shift are introduced in Eq.~(\ref{DEs}), the dynamic shift $\Delta E^{(D)}(\omega)$ is evaluated at the unperturbed Bogoliubov frequency $\epsilon_p/\hbar$, $\Delta E_0$ is the contribution to $\Delta E^{(D)}(\omega)$  from the zero mode [cf. Eqs.~(\ref{Deltap0}) and (\ref{app_DEppD})] and the final column $\Delta E$ is the total energy shift obtained as the sum of $\Delta E^{(S)}$ and $\Delta E^{(D)}$. The results for the dynamic shift are calculated using a symmetric summation and are therefore too large by about $0.01$ (see text and Fig.~\ref{fig:Econv201}). The values of $\Delta E$ in the final column should therefore be reduced by this amount, with the result that the zero temperature shift is essentially zero \cite{cite:table}.}
\label{table:energies}
\end{table*}

Another interesting feature of the results in Table~\ref{table:energies} is the enormous size of the shifts $\Delta E_{4}$ and $\Delta E_{\lambda}$ at high temperature. This is actually a separate issue from the question of infrared divergence [both these terms are part of the overall static shift $\Delta E^{(S)}$] and has its origin in the large single-particle contribution to the non-condensate density $\nt(\bfr)$ which exists at finite temperature. This is proportional to (a positive power of) the reduced temperature $t = T/T_c^0$ rather than particle interactions and hence is large for $t \sim 0.9$. However, it is shown in Ref.~\cite{Morgan04} that this single-particle contribution mostly cancels between $\Delta E_{4}$ and $\Delta E_{\lambda}$, as can be seen from their differing signs in Table~\ref{table:energies}. It is the residual large size of the sum $\Delta E_{4}+\Delta E_{\lambda}\approx\Delta E^{(S)}$ which represents the infrared divergence problem and which is removed by the inclusion of the dynamic term $\Delta E^{(D)}(\omega)$. Further discussion of this point can be found in Ref.~\cite{Morgan04}.

Although the substantial cancellation between the static and dynamic terms means that convergence in the total self-energy is much better than in these individual contributions, it also means that small fractional errors in any one term can translate to large errors in the final answer, so care is required to preserve high numerical accuracy. A further difficulty in this regard comes from the fact that a number of quantities in the theory typically converge slowly as the numerical cutoff energy $E_{cut}$ is increased. A simple example occurs in the calculation of the total number of non-condensed atoms $N_{nc}$, where at high temperature most of the particles reside in non-degenerate single-particle states above the numerical cutoff. If such quantities are required (as they may be; for example we need to calculate $N_{nc}$ in order to calculate the reduced temperature $t$) it is essential to include the effect of high-energy levels above the cutoff using the semi-classical approximation described in Appendix~\ref{sec:app:semi-classical}. 

The semi-classical approximation can also be used to improve the convergence of our numerical results for the self-energy. Once $E_{cut}$ is greater than of order $10\hbar\omega_{r}$ this approximation is highly accurate for all equilibrium quantities, and hence also for the static shifts. Indeed if it is only required to calculate static terms (as in the various versions of the HFB theories for example \cite{Hutchinson00}) then rather small values of $E_{cut}$ can be used while maintaining high accuracy. Unfortunately, the semi-classical approximation for the dynamic shift $\Delta E_p^{(D)}(\omega)$ is much less accurate (at least in the simplified form in which we have implemented it, cf. Appendix~\ref{sec:app:semi-classical}). This is awkward because we are ultimately interested in the small difference between the static and dynamic shifts and must therefore calculate these terms to the same level of accuracy. One way to overcome this problem is by brute force, and we have therefore used a very large value for $E_{cut}$ in our simulations. This reduces the importance of the semi-classical terms while increasing their accuracy to roughly  $5-10\percent$ for the dynamic self-energy. Our final results are therefore converged to within about $10^{-2}\hbar\omega_r$ at the highest temperatures we consider.

However, we have also developed a more sophisticated and vastly more accurate solution to this problem which takes into account the different ways in which a numerical cutoff should be introduced into the summations defining the static and dynamic terms. All static terms involve equilibrium non-condensate mean-fields and are therefore defined by a single summation over quasiparticle modes with a summation label $i$ [cf. the time-independent limit of Eqs.~(\ref{nt_qp})-(\ref{ci})]. Numerically, this summation is carried out for all quasiparticles modes with energies below the cutoff, $\epsilon_i \leq E_{cut}$. In contrast, the dynamic terms involve fluctuations in the non-condensate mean-fields and are calculated from a double summation over quasiparticle modes with summation labels $i$ and $j$ as in Eqs.~(\ref{app_dntrrw})-(\ref{app_Deltap}). These expressions are derived by allowing the quasiparticle wavefunctions $u_i\rt$ and $v_i\rt$ to fluctuate by writing $u_i\rt = u_i(\bfr) + \delta u_i\rt$ etc and expanding the fluctuations using the static quasiparticles as a basis via $\delta u_i\rt = \sum_{j}X_{ij}(t)u_j(\bfr)$ as in Eq.~(\ref{app_Xqpexpansion}). This expansion is the origin of the second quasiparticle index $j$ in the dynamic summations.

The two summation indices therefore have different origins and should not be subject to the same numerical cutoff. The index $i$ must encompass the same states in both the static and dynamic terms as it ultimately describes which quasiparticle modes are included in the definition of the non-condensate mean-fields. The basis associated with the index $j$ must be able to describe the dynamics of all these modes correctly, especially those with energies near $E_{cut}$, and must therefore include a greater range of states since the dynamics of a mode near the cutoff may have a significant overlap with states of higher energy. We must therefore associate a different numerical cutoff with the two summation indices and take $E_{cut}^{(j)} > E_{cut}^{(i)}$. This asymmetric summation ensures that the static and dynamic terms are calculated with comparable accuracy and produces a controlled cancellation between the corresponding self-energies for a finite basis. For the results presented in this paper we have taken $E_{cut}^{(j)} =  E_{cut}^{(i)} + 20\hbar\omega_r$ which is sufficient to provide the convergence we require. 

\begin{figure}
\psfrag{xlabel}[][]{$E_{cut}/\hbar\omega_r$}
\psfrag{ylabel}[][]{$E_p/\hbar \omega_r$ ($m=2$)}
\includegraphics[width=\columnwidth]{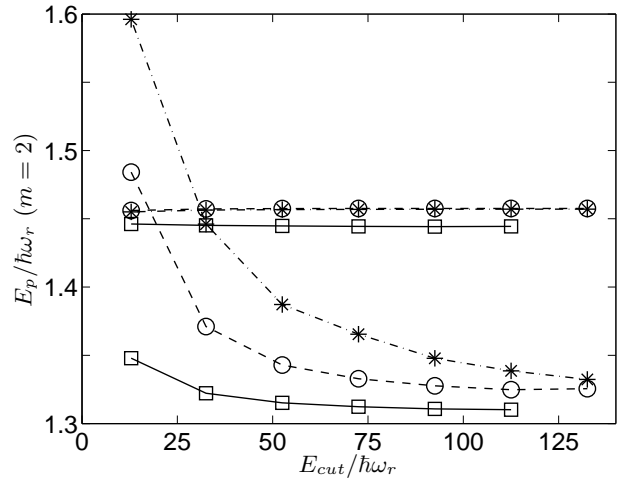}
\caption{Convergence of the quasiparticle energy $E_p$ defined in Eq.~(\ref{E_p}) for the $m = 2$ mode as a function of numerical basis energy cutoff for the parameters of the JILA experiment. Upper curves are at $T=0$, lower curves are at $T=300$nK ($t \sim 0.88$). Squares and solid line: asymmetric summation without the semi-classical approximation; circles and dashed line: symmetric summation including the semi-classical approximation; asterisk with dot-dashed line: symmetric summation without the semi-classical approximation. The lines are provided as guides for the eye.}
\label{fig:Econv201}
\end{figure}

If we are only interested in the total self-energy (rather than individual contributions to it) the use of an asymmetric summation negates the need for the semi-classical approximation and leads to much more rapid and reliable convergence, as shown in Fig.~\ref{fig:Econv201}. We also see that this new technique leads to a small but non-negligible shift in the energies at all temperatures. In particular, the shift of order $0.01\hbar\omega_r$ seen at zero-temperature in the final column of Table~\ref{table:energies} and in our earlier work \cite{Morgan03a} disappears if the dynamic self-energy is calculated using the new asymmetric summation method \cite{cite:table}. The new results are more consistent with analytical expectations \cite{Giorgini00} and are more reliable.

A difficulty with the new approach, however, is that the different ranges for the summations in the dynamic terms mean that in general the expressions for the summands can not be symmetrized with respect to the labels $i$ and $j$. This complicates the calculation, but it turns out that it is only an issue in practice for the zero temperature contributions (see Appendix~\ref{sec:app:shifts}). Nonetheless, for this reason, we have had to rederive the formulae for the dynamic terms being careful not to symmetrize with respect to the quasiparticle indices. The new expressions are given in Appendix~\ref{sec:app:shifts} and replace the corresponding ones given in \cite{Morgan04} where symmetrization was frequently used to simplify the calculation. Of course, the new results reduce to those given earlier in the case that symmetrization is allowed and in particular the expressions coincide in any exact calculation where both $E_{cut}^{(i)}$ and $E_{cut}^{(j)}$ are infinite.

The substantial cancellation of the large static and dynamic terms suggests that it may also be possible to avoid these issues by an appropriate reformulation of the theory. One possibility, first introduced by Popov, is to use density/phase variables for the perturbation theory with low-energy states. A similar approach has recently been developed and applied to quasi-condensates by Mora and Castin \cite{Mora03} and it would be an interesting subject for future work to rephrase the current calculation using this formalism.

\subsection{Finite size effects} \label{sec:finitesize}

The results in Table~\ref{table:energies} provide interesting information on the importance of various finite size contributions to the energy shift. Interactions between the equilibrium non-condensate mean-fields and the condensate affect the static condensate shape and since this provides the mean-field for the quasiparticles there is a second order effect on their energy described by $\Delta E_{\text{shape}}$. This contribution is absent in the homogeneous limit and is therefore a finite size effect, but it is clear from the table that it can be very significant. Indeed for the particular parameters of the JILA experiment, this contribution is roughly $4$-$5$ times larger than the overall energy shifts.

It is also interesting to see the importance of the contributions $\Delta E_{f}^{(S)}$ and $\Delta E_0$ which arise from a careful treatment of wavefunction orthogonality in the number conserving approach. The function $f\rt$ in the generalized GPE of Eq.~(\ref{TDGGPE}) arises specifically from the fact that the non-condensate is defined to be orthogonal to the condensate, while the contribution from $\Delta E_0$ to the dynamic shift describes the explicit effect of the two zero-energy quasiparticle modes to the dynamics of the non-condensate [see Eqs.~(\ref{Deltap0})-(\ref{app_DEppD})]. While both these contributions are small compared to the other shifts they are both comparable in size to the final answer and are therefore in principle significant. We should also point out, however, that $\Delta E_{f}^{(S)}(p)$ does not give the full contribution of the function $f\rt$ in the generalized GPE because there is also a dynamic contribution included in the quoted results for $\Delta E^{(D)}(p)$. It is also clear from the table that there is substantial cancellation between $\Delta E_{f}^{(S)}(p)$ and $\Delta E_0$ which mitigates their effect. However, we have calculated the predictions of the SOBP theory neglecting the explicit effects of both $f\rt$ and $\Delta E_0$ (but continuing to use static quasiparticles orthogonal to the condensate) and have found that the results for the energy of the $m=2$ mode clearly disagree with the measurements obtained at JILA. This calculation and its implications will be reported and discussed elsewhere \cite{Morgan03d}.

This raises an interesting and important question for future work which is to what extent the effects of these terms are reproduced in a broken symmetry description in which the quasiparticle wavefunctions are not orthogonal to the condensate, the function $f\rt$ does not appear and the quasiparticle description has to be supplemented with the `missing eigenvector' if it is to form a basis. We presume that a sufficiently careful treatment of these issues should be equivalent (for large condensates) to the results given here but the relative size of these terms in our calculation for the JILA experiment shows that such issues should not simply be ignored.

\subsection{Extracting energies and decay rates from the response function} \label{sec:fits}

\begin{figure}
\includegraphics[width=\columnwidth]{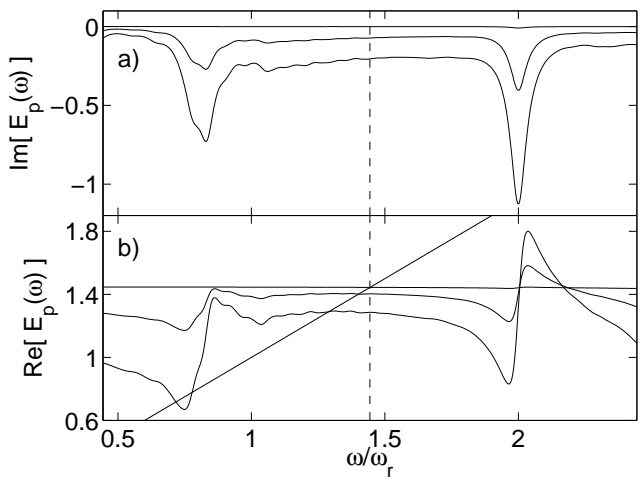}
\caption{Real and imaginary parts of the frequency-dependent quasiparticle energy $E_p(\omega)$ of Eq.~(\ref{Epw}) for the $m = 2$ mode and reduced temperatures $t=0$ (top line), $0.65$ (middle line), and $0.9$ (bottom line). a) $\mbox{Im}[E_p(\omega)]$, b) $\mbox{Re}[E_p(\omega)]$, both in units of $\hbar\omega_r$. In b) the diagonal is the line $E = \hbar\omega$ and the intersections with the curves give the poles of the resolvent $E_p$ of Eq.~(\ref{E_p}). The corresponding values of $\Gamma_p$ can then be read off from a) using Eq.~(\ref{Gamma_p}). The vertical dashed line gives the position of the unperturbed quasiparticle resonance $\epsilon_p/\hbar\omega_r = 1.44$ to 3SF.}
\label{fig:Ew201}
\end{figure}

If the self-energy $\Sigma_p(\omega+i\gamma)$ and the driving of the condensate by the thermal cloud [described by $\Delta P_{p0}^{(D)}(\omega+i\gamma)/P_{p0}(\omega)$] are roughly independent of frequency, the energy shift can be calculated straightforwardly from the poles of $\mathcal{G}_p(\omega+i\gamma)$ by finding the solutions to
\beq
E_p = \hbar \omega_p = \mbox{Re}\bigl [ E_p(\omega_p+i\gamma) \bigr ],
\label{E_p}
\eeq
with $E_p(\omega) = \epsilon_p + \Sigma_p(\omega)$ as in Eq.~(\ref{Epw}). The corresponding decay rate is then given by
\beq
\Gamma_p = -\mbox{Im}\bigl [E_p(\omega_p+i\gamma) \bigr ]/\hbar.
\label{Gamma_p}
\eeq
An example of this procedure is shown in Fig.~\ref{fig:Ew201} for the $m=2$ mode in the JILA experiment. As can be seen the self-energy is relatively smooth near the unperturbed quasiparticle resonance but has significant structure further out.

The situation of a smooth self-energy arises when an excitation couples to a continuum of decay channels, as in the homogeneous limit, and leads to resolvents $\mathcal{G}_p$ and $\mathcal{R}_p$ which are lorentzians. For a finite system, however, $\Sigma_p(\omega+i\gamma)$ depends on frequency, and both $\mathcal{G}_p$ and $\mathcal{R}_p$ can differ significantly from perfect lorentzians. In this case the line shape depends on the details of the system, and we have to extract energies and decay rates by fitting the response to a suitable function. When the spectrum is strongly distorted (as it can be at high temperature for example), the results obtained from these fits can be sensitive to the exact fitting procedure and this ultimately puts error bars on the theoretical predictions. We have therefore implemented two fitting procedures to determine the energies and decay rates: in our earlier work we used a complex lorentzian to fit the full response in the frequency domain \cite{Morgan03a}, while in the present paper we mimic the experimental procedure exactly and fit a decaying sinusoid to the response in the time-domain.

We find that at low temperatures both methods produce excellent fits and the energies and decay rates can be extracted straightforwardly. At higher temperatures, however, the spectra develop noticeable non-lorentzian structure (see Figs.~\ref{fig:EGRw201} and \ref{fig:EGRw001}) and it becomes more difficult to obtain reliable fits. This is not surprising when one considers that a fit is an attempt to model a complex spectrum with very few parameters which is difficult to do reliably when the spectrum does not have the assumed form. The problem is exacerbated from the fact that the intrinsic width of the spectra at high temperature is of order $0.2\hbar\omega_r$ (cf. Fig.~\ref{fig:gvtm2m0}) while we would like to determine the central frequency to an accuracy roughly an order of magnitude better than this.

These problems can be overcome to some extent by including the spectrum of the perturbation $P_{p0}(\omega)$ as a known weight function in the fit. This ensures that only the experimentally relevant range of frequencies are included and has the effect of suppressing some of the non-lorentzian structure in the wings of the distribution. Specifically, we fit the spectra in the frequency domain to the function
\beq
f(\omega) = P_{p0}(\omega)\bigl [ \frac{A}{\hbar\omega-E_p+i\Gamma_p} + C \bigr ],
\label{wfit_formula}
\eeq
where the fit parameters are the real constants $E_p$ and $\Gamma_p$ and the complex constants $A$ and $C$ [$C$ accounts for any linear frequency dependence of $\Delta P_{p0}^{(D)}(\omega)$]. The experimental resolution $\gamma$ is subtracted from the fitted width parameter $\Gamma_p$ to give the intrinsic width. The weight function $P_{p0}(\omega)$ is calculated using the functional form given in Eq.~(\ref{perturbation}) and has the frequency dependence
\beq
P_{p0}(\omega) \propto \frac{\sin[(\omega-\omega_d)t_d]}{(\omega-\omega_d)t_d}.
\eeq
The results are not particularly sensitive to the choice of the central drive frequency $\omega_d$, which was chosen in the experiment to maximise the observed response. In our calculations we chose it to be equal to the unperturbed Bogoliubov energy $\epsilon_p$ at $T=0$, while at higher temperatures we take it to have the value of $E_p$ found at the closest lower temperature for which we have data. In this way the trend of the central drive frequency broadly follows the trend of the energy shift.

For the time-domain fits, we take the Fourier transform of the response functions (including the factor $P_{p0}(\omega)$ as above) and fit the real part (which describes the real condensate density oscillations) to a decaying sinusoid of the form $Ae^{-\Gamma_p t}\sin(E_p t + \phi) + B$ where the fit parameters are all real. The times $t$ are chosen in the range $t_d < t < t_{obs}$ where $t_{obs}$ is the experimental observation time, which is $34$ms in the JILA experiment \cite{Jin97}. The resolution parameter $\gamma$ (which is no longer required) is removed by multiplying the oscillations by $e^{+\gamma t}$ before fitting. The results from this procedure are in complete agreement with the fits in the frequency domain at low temperatures where the spectra are well described by lorentzians, but at high temperatures there are differences of the order of a few $10^{-2}\hbar\omega_{r}$. This uncertainty in how to extract energies and decay rates from non-lorentzian spectra therefore represents the theoretical uncertainty in the predictions.

Looking ahead to the results shown in Fig.~\ref{fig:Evtm2m0}, we see that the energy $E_p$ extracted from assuming a frequency-independent self-energy is clearly in better agreement with experiment for the $m=2$ mode at the highest temperature than the values extracted from the fits. Inspection of the shape of the frequency-dependent quasiparticle energy for this case, given in Fig.~\ref{fig:Ew201}, shows that the values of $E_p$ are indeed representative of the typical values in the frequency range of interest. Given the arduous nature of the calculation, it is therefore somewhat frustrating that the agreement with experiment should be substantially worsened by the fitting procedure. Nonetheless, fitting the condensate density oscillations in the time-domain mimics the experimental procedure and is therefore presumably the appropriate method to use when the spectrum has non-lorentzian structure.

\section{Numerical results} \label{sec:numerical:results}

In this section we present a detailed analysis of the SOBP predictions for the 1997 JILA experiment \cite{Jin97}. In particular, we show the functional form of the response functions including and excluding the effect of direct excitation of the non-condensate and discuss the underlying physics. We then present results for the energies and decay rates of the lowest energy $m=2$ and $m=0$ modes, corresponding to the quantum numbers $(m,p,n) = (2,0,1)$ and $(0,0,1)$ respectively. In Sec.~\ref{sec:results:N0} we provide a detailed analysis of the experimental results for condensate population $N_0$ and temperature $T$ to determine the appropriate values to use in our simulations and the uncertainties which variations in these quantities can be expected to produce in our results (with the exception of the analysis in this section we have consistently taken $N_0 = 6000$). We also present results for the axial and radial dipole modes (quantum numbers $(m,p,n) = (0,1,1)$ and $(\pm1,0,1)$ respectively) and show that these are obtained correctly in the SOBP theory provided that excitation of the non-condensate is included. Finally, we discuss the relative phase of the condensate and non-condensate density oscillations.

\subsection{Response functions} \label{sec:response_functions}

In this section we consider the functional form of the resolvents $\mathcal{G}_p(\omega)$ and $\mathcal{R}_p(\omega)$ defined in Eqs.~(\ref{G}) and (\ref{R}) respectively for the lowest energy $m=2$ and $m=0$ modes. For the $m=2$ mode, symmetry considerations mean that there are no fluctuations in the condensate population ($\delta N_0 (\omega)$=0) and these quantities are directly proportional to the condensate density response in the frequency domain. For the $m=0$ mode, however, $\delta N_0 (\omega)$ is non-zero and should be included in the analysis of the density response. This can be done as described in Appendix~\ref{sec:app:condDensflucs}, but has very little effect on the detailed results. For the purposes of the present discussion, the difference is not substantial enough to merit complicating the issue further and we will therefore treat $\mathcal{G}_p(\omega)$ and $\mathcal{R}_p(\omega)$ as giving the density response directly for both modes. The only exception to this comes in Sec.~\ref{sec:phase} where we consider the relative phase of the condensate--non-condensate oscillations and include the effect of $\delta N_0 (\omega)$ for the $m=0$ mode.

\begin{figure}
\includegraphics[width=\columnwidth]{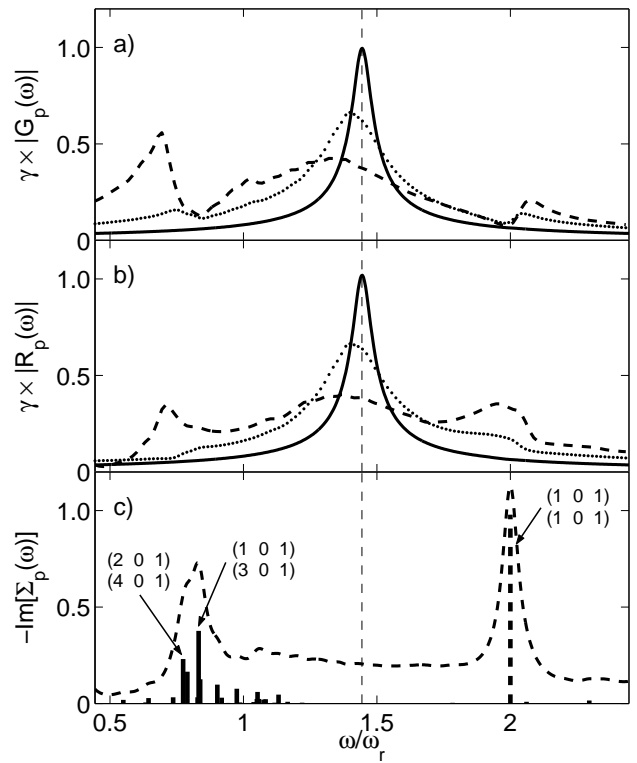}
\caption{a) $\gamma \times |\mathcal{G}_p(\omega)|$, b) $\gamma \times |\mathcal{R}_p(\omega)|$: modulus of resolvent as a function of frequency for the $m = 2$ mode [$(m,p,n) = (2,0,1)$] and $t=0$ (solid), $t=0.65$ (dotted, $\times 2$), and $t=0.9$ (dashed, $\times 3$). c) $-\mbox{Im}[\Sigma_p(\omega)]$ at $t = 0.9$ (dashed curve) and a few contributing Landau and Beliaev processes (solid and thick-dashed vertical lines respectively). These processes are drawn at their resonance frequencies $\omega_{ij}$ with a height corresponding to their amplitude in the self-energy. The three largest contributions are labelled with the $(m,p,n)$ quantum numbers for the two quasiparticle modes involved (the top label applies to the mode with the lower energy). The thin, vertical, dashed line indicates the position of the unperturbed Bogoliubov resonance, $\epsilon_p = 1.44\hbar\omega_r$ to 3SF.}
\label{fig:EGRw201}
\end{figure}
\begin{figure}
\includegraphics[width=\columnwidth]{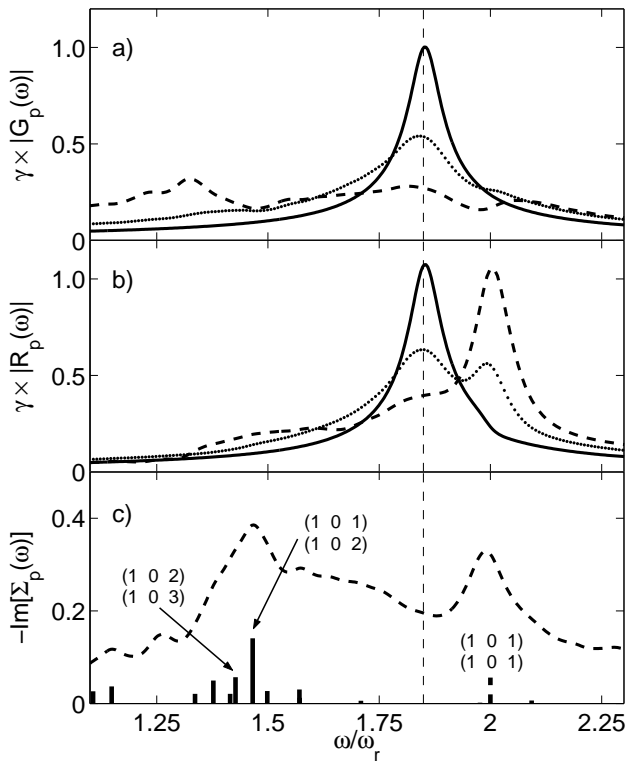}
\caption{Plots as in Fig.~\ref{fig:EGRw201} but for the $m=0$ mode [$(m,p,n) = (0,0,1)$]. The curves in a) and b) now correspond to: $t=0$ (solid), $t=0.65$ (dotted, $\times 1.5$), and $t=0.9$ (dashed, $\times 2$). The unperturbed Bogoliubov resonance is at $\epsilon_p = 1.85\hbar\omega_r$ to 3SF.}
\label{fig:EGRw001}
\end{figure}

The absolute values of these response functions are plotted in Figs.~\ref{fig:EGRw201} and \ref{fig:EGRw001} for the $m=2$ and $m=0$ modes respectively as functions of frequency $\omega$ for a range of reduced temperatures. In each case, the upper panel (a) shows $\mathcal{G}_p(\omega)$, which neglects direct driving of the non-condensate by the perturbation, while the middle panel (b) shows the full response $\mathcal{R}_p(\omega)$ where this process is included. The lower panel (c) shows the contribution to the self-energy from a subset of low-energy Landau and Beliaev processes and is included to aid analysis of the response function plots.

We focus first on the results for $\mathcal{G}_p(\omega)$. As can be seen, at $t=0$ the response of both the $m=2$ and $m=0$ modes is a lorentzian positioned almost exactly at the frequency of the unperturbed Bogoliubov energy. In fact, we find that the shift in the frequencies at zero temperature is essentially zero for both modes (see Fig.~\ref{fig:Evtm2m0}), in contrast with our earlier calculations where shifts of order $2\times 10^{-2}\hbar\omega_r$ were found \cite{Morgan03a}. The new results are consistent with the analytical results of Giorgini \cite{Giorgini00} obtained in the Thomas-Fermi regime, and are expected to be more accurate because they incorporate the asymmetric summation method described in Sec.~\ref{sec:convergence} which has a substantial effect at zero-temperature, as shown in Fig~\ref{fig:Econv201}. The width of the response functions at zero temperature is entirely due to the parameter $\gamma$ introduced to model the finite experimental resolution. As the temperature increases, however, the response develops an intrinsic width and there is a noticeable downwards shift in the position of the peak with a concomitant decrease in height.

At the highest temperatures, considerable structure is observable in the wings of the response, which is no longer a simple lorentzian. The physical processes responsible for this structure can be identified using the self-energy plots of Figs.~\ref{fig:EGRw201}c and \ref{fig:EGRw001}c. Here we show the negative of the imaginary part of the self-energy at a high temperature $t = 0.9$ (the dashed curve) as well as vertical bars whose height and position indicate the contribution from individual Landau and Beliaev processes. As Eq.~(\ref{dD}) demonstrates, the imaginary part of a dynamic term is formed from the combined effect of many such processes, each broadened by the resolution parameter $\gamma$. Where a large number of these resonances overlap, the self-energy is smooth but if a few large resonances dominate at some frequency then sharp features are observed leading to non-lorentzian structure in the resolvent. This structure generally occurs at a frequency slightly further from the Bogoliubov energy $\epsilon_p$ than the features in the self-energy because of level repulsion (typical of a second order perturbation calculation) between these processes and the central resonance.

For the $m=2$ mode, Figs.~\ref{fig:EGRw201}a and \ref{fig:EGRw201}c show that the structure in $\mathcal{G}_p(\omega)$ around $\omega/\omega_r = 0.7$ is a consequence of a few Landau processes of the form $(2,0,1) + (m,0,n) \rightarrow (m+2,0,n)$ for small $m$, while that near $\omega/\omega_r \sim 2$ is a result of a single strong Beliaev decay into two \textit{x-y} dipole modes, $(2,0,1) \rightarrow (1,0,1)+(1,0,1)$. For the $m=0$ mode, the structure around $\omega/\omega_r \sim 1.3$ in Fig.~\ref{fig:EGRw001}a is the result of a few Landau processes of the form $(0,0,1)+(m,p,n) \rightarrow (m,p,n+1)$ for small values of $m$. For both modes, the processes described involve low-energy collective excitations of the thermal cloud, with the result that they have quite large matrix elements and hence can be significant at finite temperature when the associated rates are strongly enhanced by Bose stimulation. For the TOP trap geometry, these processes occur at frequencies quite a long way from the principal resonance, but the large width of the response at finite temperature means that they can still cause a significant distortion of the spectrum, even to the extent that the greatest response no longer occurs in the vicinity of the original Bogoliubov mode. The effect of these processes could be increased by tuning the trap geometry to shift them closer to resonance, although a full discussion of their importance requires the inclusion of direct excitation of the non-condensate by the perturbation, which we now consider.

The effect of including direct driving of the non-condensate is shown in Figs.~\ref{fig:EGRw201}b and \ref{fig:EGRw001}b, where the full response function $\mathcal{R}_p(\omega)$ is plotted. For the $m=2$ mode this process only affects the wings of the response, which is otherwise qualitatively the same as if it is neglected. The main change is an enhancement of the structure around $\omega/\omega_r \sim 2$, which is again a consequence of a single Beliaev process in the thermal cloud involving the excitation of two $(1,0,1)$ dipole modes. For the $m=0$ mode, however, there is a dramatic change in the form of the response function at high temperature, with a growing peak at $\omega/\omega_r = 2$ which eventually dominates the spectrum. In this case, the change in the response is due, not to a single Beliaev process, but rather to a large number of weak Landau processes. The external perturbation is proportional to $r^2$ and hence couples strongly to high-energy, single-particle modes in the thermal cloud. Since these modes are weakly-interacting, there are a large number of Landau processes with frequency differences near the ideal gas value of $2\omega_r$. When the temperature is high enough that these modes are significantly populated, the non-condensate has a large response at this frequency, which it can transfer to the condensate via their dynamical coupling. This process ultimately dominates the condensate response, with the result that at high temperature it is more likely to be excited indirectly via the thermal cloud than directly via the perturbation. This provides the microscopic explanation for the strong upward shift in the excitation energy of the $m=0$ mode at $t\sim 0.6$ observed in the 1997 JILA experiment (see Fig.~\ref{fig:Evtm2m0}) \cite{Jin97,Morgan03a}.

\begin{figure}
\psfrag{ylabel}[][]{$|\tilde{\mathcal{G}}_p(\omega)/\mathcal{G}_p(\omega)|$}
\includegraphics[width=\columnwidth]{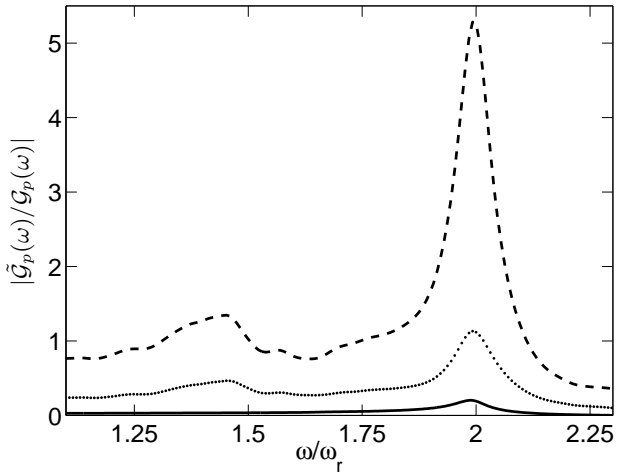}
\caption{Plot of the absolute value of $\tilde{\mathcal{G}}_p(\omega)/\mathcal{G}_p(\omega)$ as a function of frequency for the $m=0$ mode. As shown by Eq.~(\ref{Gt}), this gives the ratio of indirect condensate excitation via the thermal cloud to its direct excitation by the external perturbation. Curves are shown for $t=0$ (solid), $t=0.65$ (dotted), and $t=0.9$ (dashed) as in Fig.~\ref{fig:EGRw001}.}
\label{fig:GtGwm0}
\end{figure}

The importance of including the direct excitation of the non-condensate by the external perturbation is perhaps shown most clearly in Fig.~\ref{fig:GtGwm0} where we plot $\tilde{\mathcal{G}}_p(\omega)/\mathcal{G}_p(\omega)$ which describes the ratio of the indirect excitation of the condensate via the thermal cloud to its direct excitation by the external perturbation [cf. Eq.~(\ref{Gt})]. At low temperatures and for most frequencies, this ratio is much less than one indicating that the condensate is mostly driven by the external perturbation. However, there is also a peak at $\omega = 2\omega_r$ which grows with temperature and at $t=0.9$ we find that the condensate is driven roughly five times as strongly via the thermal cloud as it is directly via the perturbation.

We should note that whereas the results for $\mathcal{G}_p(\omega)$ are fundamental, depending on the intrinsic couplings between the condensate and thermal cloud, the response function $\mathcal{R}_p(\omega)$ depends on the assumed form of the external perturbation and as such will differ in detail from one experiment to another. This provides a handle to explore the relative importance of the two mechanisms for exciting the condensate. For example, if the perturbation is chosen to be spatially localized around the centre of the trap (rather than the $r^2$ form assumed here) then the effect of excitation of the thermal cloud will be greatly reduced and the condensate response should be well described by $\mathcal{G}_p(\omega)$ alone. It would be interesting therefore if the response of the $m=0$ mode was remeasured using perturbations of different spatial form so that the downward shift for this mode predicted in the absence of thermal cloud driving can also be observed (see Fig.~\ref{fig:Evtm2m0}).

\subsection{Energy shifts and decay rates} \label{sec:energies}

\begin{figure}
\psfrag{xlabel}[][]{$t = T/T_c^0$}
\psfrag{ylabela}[][]{$E/\hbar \omega_r$ ($m=0$)}
\psfrag{ylabelb}[][]{$E/\hbar \omega_r$ ($m=2$)}
\includegraphics[width=\columnwidth]{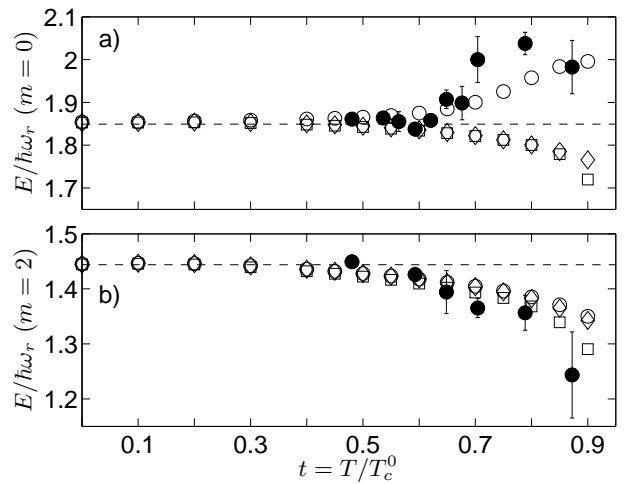}
\caption{Ab initio theoretical excitation energies $E$ (open symbols) compared with experiment (filled circles with error bars) for (a) the $m = 0$ and (b) the $m = 2$ mode. Diamonds neglect direct thermal driving (${\cal G}_p$), open circles include it (${\cal R}_p$) and squares give $E_p$ of Eq.~(\ref{E_p}). The dashed line is the Bogoliubov energy $\epsilon_p$. Differences between diamonds and squares are due to non-Lorentzian structure in ${\cal G}_p$. There are no free parameters in the theoretical results.}
\label{fig:Evtm2m0}
\end{figure}

\begin{figure}
\psfrag{xlabel}[][]{$t = T/T_c^0$}
\psfrag{ylabela}[][]{$\Gamma/\omega_r$ ($m = 0$)}
\psfrag{ylabelb}[][]{$\Gamma/\omega_r$ ($m = 2$)}
\includegraphics[width=\columnwidth]{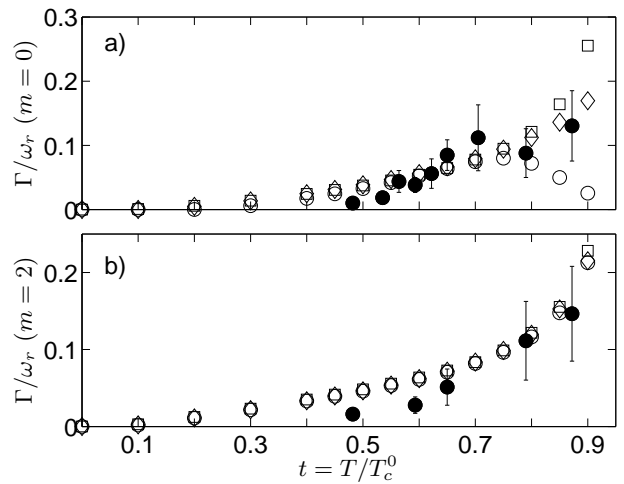}
\caption{Theoretical decay rates ($\Gamma$) compared with experiment for (a) the $m = 0$ mode and (b) the $m = 2$ mode. Symbols as in Fig.~\ref{fig:Evtm2m0}.}
\label{fig:gvtm2m0}
\end{figure}

We extract energies and decay rates from the resolvents given in Figs.~\ref{fig:EGRw201} and \ref{fig:EGRw001} by finding poles of the self-energy and by using fits to the oscillations in the time-domain, as described in Sec.~\ref{sec:fits}. The results of these calculations for the $m=0$ and $m=2$ modes are shown in Figs.~\ref{fig:Evtm2m0} and \ref{fig:gvtm2m0}, and are essentially the same as those given in Ref.~\cite{Morgan03a}. There are two minor differences compared with our earlier work, however. The first comes from the use of the asymmetric summation method which shifts the energies down slightly at all temperatures and reduces the zero-temperature shifts to almost zero. The second comes from fitting the spectra in the time-domain rather than the frequency domain which makes the upward shift of the $m=0$ mode at $t\sim 0.6$ somewhat less abrupt when direct driving of the thermal cloud is included. However, all our earlier conclusions remain unchanged, and as we commented in Sec.~\ref{sec:fits} the difficulty in extracting meaningful fits from non-lorentzian spectra represents the theoretical uncertainty in the final predictions. 

It is perhaps worth noting that a critical analysis of Fig.~\ref{fig:Evtm2m0}b suggests that the theory slightly underestimates the shifts seen in the experiment at high temperature for the $m=2$ mode. Although the error bar on the final point is rather large in this case, this may be an indication that the experiment is seeing effects beyond the SOBP theory. Recently, Liu \textit{et al.} \cite{Liu04} have included self-consistent effects of the non-condensate mean-field dynamics for a spherical trap geometry and shown that this has the effect of enhancing the shifts predicted by the SOBP theory for the monopole mode. It would therefore be interesting to apply this approach to the JILA experiment to see if it improves agreement with experiment for the $m=2$ mode at high temperature. However, the calculation of \cite{Liu04} neglects collisional effects in the non-condensate which may also be significant in a systematic extension of the SOBP theory to next order in the dilute gas parameter. These effects are included in the recent theories of Walser \textit{et al.} \cite{Walser99,Walser00}, Wachter \textit{et al.} \cite{Wachter02} and Proukakis \cite{Proukakis01}, although to the best of our knowledge these approaches have also not been applied to anisotropic geometries.

\subsection{Effect of changes in condensate population} \label{sec:results:N0}

\begin{figure}
\includegraphics[width=\columnwidth]{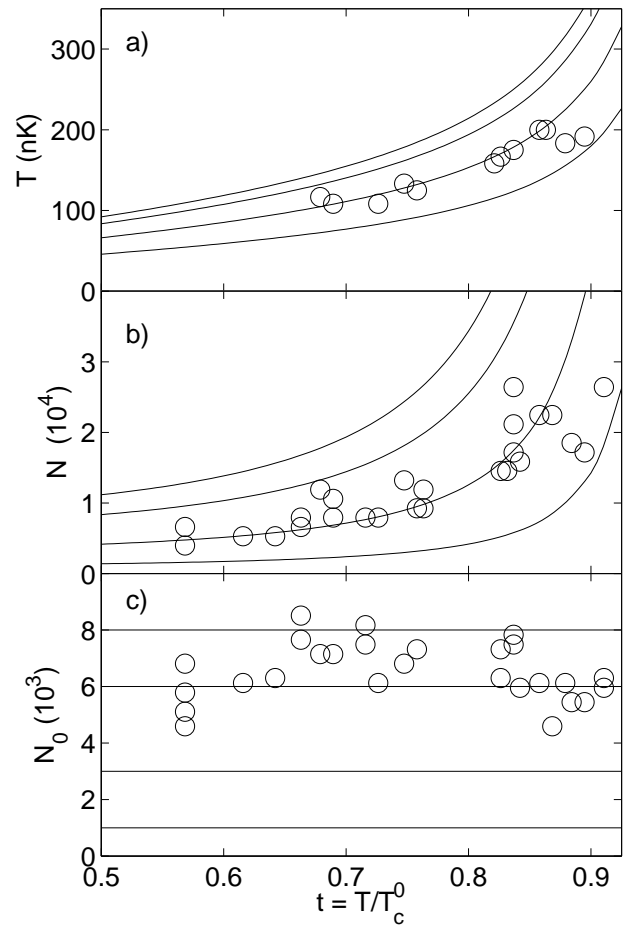}
\caption{Bogoliubov theory (solid lines) compared with experiment (open circles) for (a) $T$ in nanokelvin vs $t$, (b) $N$ vs $t$, and (b) $N_0$ vs $t$. Solid lines correspond to $N_0 = 8000$, $6000$, $3000$, and $1000$ from top to bottom in each case.}
\label{fig:TNN0vst}
\end{figure}

For a fixed trap geometry and atomic species, the main input parameters to the numerical calculations are the condensate population $N_0$ and the absolute temperature $T$ in nanokelvin. The relevant experimental data for these quantities is given in Fig.~1 of Ref.~\cite{Jin97} and is reproduced in Fig.~\ref{fig:TNN0vst}, where $N_0$, $T$ and the total atom number $N$ are plotted as functions of reduced temperature $t$ \cite{errorbars}. In order to verify the correct inputs for our calculation, we have used the ordinary Bogoliubov theory with $N_0$ and $T$ as input to calculate the total atom number $N$ from the number of non-condensed atoms $N_{nc}(T,N_0) = \intdr \nt (\bfr)$ via $N(T,N_0) = N_0+N_{nc}(T,N_0)$. This is then used to calculate the reduced temperature $t$ using Eq.~(\ref{Tc0}) to obtain the ideal gas critical temperature $T_c^0$. The results of these calculations for four values of $N_0$ and a range of temperatures are shown in Fig.~\ref{fig:TNN0vst} as solid lines.

This calculation shows that while the experimental data for $N_0$ is consistent with a condensate population in the region of $N_0 = 6000$ or greater for $t<0.9$, the results for $N$ and $T$ are more consistent with $N_0 = 3000$. Since the Bogoliubov theory should adequately describe the thermodynamics of the experiment, we conclude that there may be substantial error (possibly systematic) in some of the experimental data. We stress that these numerical calculations come from a full implementation of the ordinary Bogoliubov theory for the anisotropic trap geometry and finite particle number of the experiment and the convergence on all numerical quantities shown is of order a part in $10^5$.

\begin{figure}
\psfrag{xlabel}[][]{$t = T/T_c^0$}
\psfrag{ylabela}[][]{$E/\hbar \omega_r$ ($m=0$)}
\psfrag{ylabelb}[][]{$E/\hbar \omega_r$ ($m=2$)}
\includegraphics[width=\columnwidth]{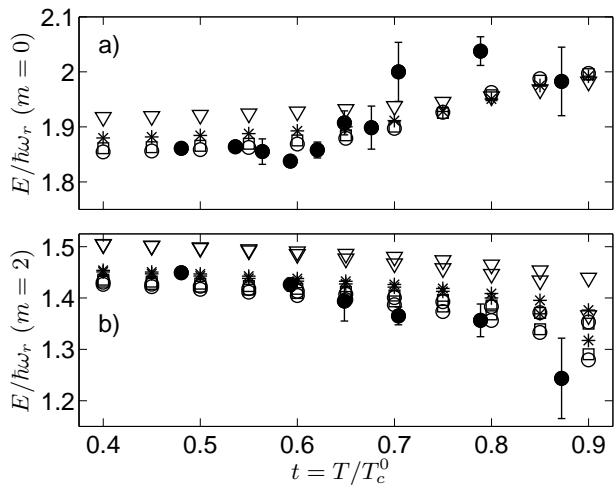}
\caption{Quasiparticle energies for a) the $m=0$ mode and b) the $m=2$ mode as a function of reduced temperature $t$ for $N_0 = 1000$ (triangles), $N_0 = 3000$ (stars), $6000$ (squares) and $8000$ (circles). The results for $m=0$ and the upper set of points for $m=2$ come from fits to the resolvent $\mathcal{R}_p$ in the time domain, while the lower set of results for $m=2$ correspond to the value of $E_p$ from Eq.~(\ref{E_p}). The difficulty in extracting meaningful energies from non-lorentzian spectra is visible by the large difference between these two sets of results for $t>0.8$.}
\label{fig:EtN0_m2m0}
\end{figure}

We have assumed that the data for the condensate population is probably the most reliable and for this reason we have used a value of $N_0=6000$ for most of our calculations and for the results reported in Ref.~\cite{Morgan03a}. Given the uncertainty in $N_0$, however, it is important to calculate what effect this has on our earlier predictions. In Fig.~\ref{fig:EtN0_m2m0} we plot the energies obtained from the SOBP theory for both the $m=2$ and $m=0$ modes for $N_0 = 1000, 3000, 6000$ and $8000$, which cover the range of possible values in the experiment. As can be seen the results are relatively insensitive to the value of $N_0$, although the case $N_0 = 1000$ is clearly excluded. Indeed, at high temperature the difficulty in extracting a meaningful energy from a non-lorentzian spectrum leads to a similar uncertainty in the prediction as variation of the condensate population within the relevant range. We conclude that the uncertainty in the relevant values of $N_0$ and $T$ to use in our simulations has little effect on our overall results, although clearly it would be advantageous to have new experimental results where this uncertainty was removed. 

\subsection{Dipole modes} \label{sec:dipole}

In a trapped system at finite temperature both the condensed and non-condensed atoms can undergo centre-of-mass oscillations, corresponding to the excitation of the dipole modes of the system. If there is relative motion of the two clouds then these oscillations will ultimately be damped \cite{StamperKurn98}. In a harmonic potential, however, the generalized Kohn theorem \cite{Dobson94} shows that in-phase oscillations are an exact solution of the full equations of motion and are completely decoupled from any internal dynamics of the clouds, independent of both temperature and particle interactions. As a consequence we have the exact result that the system has undamped dipole oscillations with energies corresponding exactly to the principal trap frequencies. Whether or not these modes are obtained correctly is therefore an important test of any theoretical description of a condensed system at finite temperature.

In Ref.~\cite{Morgan04} we showed that the SOBP theory is consistent with the generalized Kohn theorem for the dipole modes to within the small parameter of the theory. This is only the case, however, if the effect of the external perturbation on the non-condensate dynamics is included. This makes sense because it is clearly not possible to describe an in-phase oscillation of the condensate and non-condensate if the force which generates the motion is assumed at the outset to act only on the condensate. 

\begin{figure}
\psfrag{xlabel}[][]{$\omega/\omega_z$}
\psfrag{ylabela}[][]{$\gamma \times |\mathcal{G}_p(\omega)|$}
\psfrag{ylabelb}[][]{$\gamma \times |\tilde{\mathcal{G}}_p(\omega)|$}
\psfrag{ylabelc}[][]{$\gamma \times |\mathcal{R}_p(\omega)|$}
\includegraphics[width=\columnwidth]{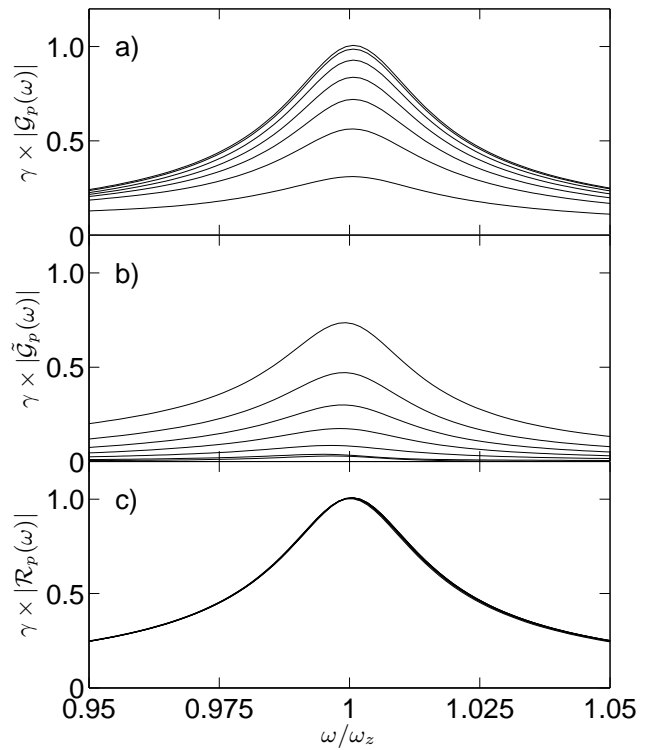}
\caption{Plot of the absolute value of the resolvents (a) $\mathcal{G}_p(\omega)$, (b) $\tilde{\mathcal{G}}_p(\omega)$ and (c) $\mathcal{R}_p(\omega)$ as a function of angular frequency $\omega$ for the dipole oscillation along the \textit{z}-axis, $(m,p,n) = (0,1,1)$. In all cases the spectra are shown for reduced temperatures $t = 0, 0.15, 0.3, 0.45, 0.6, 0.75$, and $0.9$. In (a), temperature increases from top to bottom, in (b) it increases from bottom to top, while in (c) the various curves are indistinguishable on this scale.}
\label{fig:Kohn}
\end{figure}
Our numerical results for the axial dipole oscillation are shown in Fig.~\ref{fig:Kohn} where we plot the magnitude of the response functions $\mathcal{G}_p(\omega)$, $\tilde{\mathcal{G}}_p(\omega)$ and $\mathcal{R}_p(\omega)$ for a range of temperatures. In Fig.~\ref{fig:Kohn}a, the effect of direct excitation of the non-condensate is excluded and so the dipole mode is not obtained exactly. In this case, the perturbation moves the condensate relative to the centre of the non-condensed atoms with the result that the oscillations are damped. This is visible as an increase in the width of the response function and a decrease in its peak height as the temperature increases (there is no significant change in the position of the peak frequency). The damping is very light compared to the other low-energy modes, however: at $t \sim 0.9$ we have $\Gamma_p \sim 0.03\omega_r$, which is $20\percent$ or less of the decay rate for the $m=0$ and $m=2$ modes shown in Fig.~\ref{fig:gvtm2m0}.

It is also interesting to consider the condensate response in the case where it is driven \textit{only} via the thermal cloud, corresponding to the experiment of Ref.~\cite{StamperKurn98}. This is described by the response function $\tilde{\mathcal{G}}_p(\omega)$ defined in Eq.~(\ref{Gt}) and is shown in Fig.~\ref{fig:Kohn}b for the case that $P\rw \propto z$, which is the linear form required to excite a pure axial dipole oscillation. In this case the amplitude of the response increases with temperature as the size and importance of the thermal cloud increases. Although not obvious from the figure, the width of this response function is also slightly narrower than the parameter $\gamma$ we introduced into the resolvents to model the experimental resolution. This means that when we transform $\tilde{\mathcal{G}}_p(\omega)$ to the time-domain and remove this (artificial) decay by multiplying by $e^{+\gamma t}$ we see a clear growth in the amplitude of the condensate density oscillations with time. This is in contrast to the case for Fig.~\ref{fig:Kohn}a where the oscillations decay with time and conforms precisely with our expectations: In this case the perturbation causes the thermal cloud to oscillate through the condensate resulting in a transfer of energy and momentum and hence to condensate oscillations which initially grow with time.

The full condensate density response is obtained by adding $\mathcal{G}_p(\omega)$ and $\tilde{\mathcal{G}}_p(\omega)$ to obtain $\mathcal{R}_p(\omega)$, which is shown in Fig.~\ref{fig:Kohn}c. In this case the curves for the various temperatures collapse onto each other, and to high accuracy the response is a pure lorentzian centred on the axial trap frequency. As expected, there is no damping of the oscillation and the width of the response shown simply corresponds to our resolution parameter $\gamma$. Extracting the energy and decay rate from the spectra, we find that the peak of the lorentzian is at the axial trap frequency $\omega_z$ to within $0.1\percent$ for all temperatures, which is of the order of our numerical accuracy, while the intrinsic decay rate is indistinguishable from zero. This result confirms that this dipole mode is obtained correctly in the SOBP theory and acts as an important check on our numerical method.

We have also calculated the spectra for the dipole oscillations in the \textit{x-y} plane [quantum numbers $(m,p,n) = (\pm1,0,1)$] with very similar results. If direct excitation of the non-condensate is included, we find that the mode frequency is obtained correctly to within $0.1\percent$ for $t<0.6$, although the error rises to nearly $1\percent$ at $t = 0.9$. The increased error in this case may be due to the approximation that only the positive frequency pole of the condensate response is relevant in the calculation \cite{Morgan04}. In reality, the full response also has a contribution from a pole at $\omega = -\omega_r$ and this is a factor of $\sqrt{8}$ closer to the positive frequencies of interest for radial oscillations than it is for axial oscillations in a TOP trap geometry.

Finally, we note that the use of the asymmetric summation technique described in Sec.~\ref{sec:convergence} greatly improves the accuracy of our calculation of the dipole modes, especially at zero temperature. If a symmetric summation is used instead, we find a systematic upwards shift in the zero-temperature frequency of both the radial and axial dipole modes of order $2\times10^{-2}\omega_r$, representing a $2\percent$ error for the radial modes. The asymmetric summation reduces this error by an order of magnitude to within the level of our numerical accuracy. 

\subsection{Relative phase of condensate--non-condensate oscillations} \label{sec:phase}

An important issue in the study of condensate dynamics at finite temperature is the relative phase of the oscillations of the condensed and non-condensed atoms. Indeed, it has been argued by Bijlsma and Stoof \cite{Bijlsma99} and by Al Khawaja and Stoof \cite{Khawaja00} that the JILA results for the $m=0$ mode can be explained by a transition from an out-of-phase motion of the two clouds to an in-phase motion at high temperature. We have found that the SOBP calculation essentially confirms this conclusion. Combining the expression for the condensate density fluctuations in Eq.~(\ref{dnc_GGPE}) (including the explicit effect of the condensate number fluctuations $\delta N_0(\omega)$ for the $m=0$ mode) with that for the non-condensate in Eq.~(\ref{app_dntrrw}) we can straightforwardly calculate their relative phase. Of course, both these quantities are functions of position and frequency as well as temperature, so we first take a suitable moment of both oscillations as described in Eqs.~(\ref{app_GRdN0}) and (\ref{app_GRdnt}) and then evaluate their relative phase at the frequency which gives the best fit to the condensate oscillations (averaged over a region of $2\gamma$). This gives a measure of the relative phase of the two oscillations at the peak condensate response as a function of temperature.

\begin{figure}
\includegraphics[width=\columnwidth]{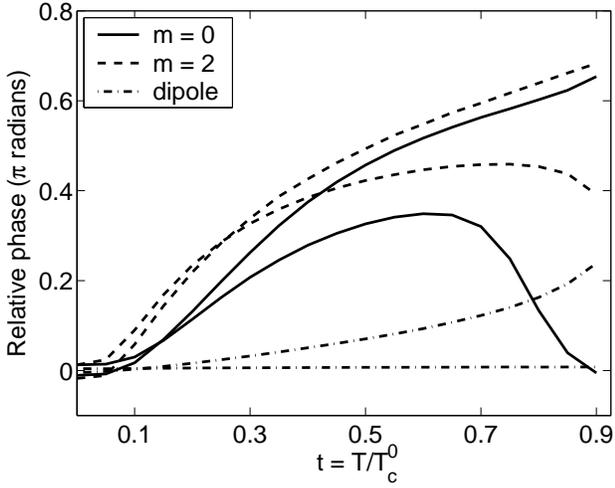}
\caption{Plot of the phase of the non-condensate density oscillations relative to those of the condensate as a function of reduced temperature for the parameters of the JILA experiment. The solid line gives the results for the $m=0$ mode, the dashed line for the $m=2$ mode and the dashed-dotted line for the axial dipole mode. In each case the upper curve neglects direct driving of the non-condensate by the external perturbation while this effect is included in the lower curve.}
\label{fig:phase}
\end{figure}

The results of this calculation for the $m=2$ and $m=0$ modes as well as the axial dipole mode are shown in Fig.~\ref{fig:phase}. As can be seen, in all cases the oscillations are in phase near $t=0$. If direct driving of the non-condensate is neglected, the oscillations become increasingly out-of-phase as the temperature rises, especially for the $m=2$ and $m=0$ modes. The results are very different when the effect of the perturbation on the non-condensate is included. In this case we see that the $m=0$ mode shows a distinct crossover to an in-phase motion at high temperatures. This is indeed what we would expect from our earlier analysis as in this case the condensate is driven predominately by the thermal cloud at a frequency above its natural Bogoliubov energy resulting in an in-phase response. A similar effect is also seen for the $m=2$ mode although it is less strong because the thermal cloud resonance at $\omega = 2\omega_r$ is much less significant in this case. It is also gratifying to see that the condensate and non-condensate oscillations for the dipole mode remain locked in-phase at all temperatures when direct driving of the thermal cloud is included. This is consistent with the Kohn theorem and with the results we obtained for the dipole modes in Sec.~\ref{sec:dipole}. 

\section{Conclusions}

In conclusion, we have presented a detailed analysis of the measurements of condensate excitations made at JILA in 1997 \cite{Jin97} using a theory of the linear response of Bose-Einstein condensates at finite temperature that we have recently derived. We have shown the importance of including the direct effect of the external perturbation on the non-condensate dynamics and in particular we have demonstrated unambiguously that this provides the explanation for the anomalous behaviour of the $m=0$ mode observed at JILA and that it is necessary for a correct description of the dipole modes. A major issue in the numerical implementation of the theory is achieving convergence of the results and we describe a novel asymmetric summation scheme we have developed which solves this problem. This makes it feasible to apply our theoretical formalism to a wide variety of more recent experiments involving larger condensate populations and more highly anisotropic traps.

\begin{acknowledgments}
It is a pleasure to thank Keith Burnett, Matthew Davis, Simon Gardiner, David Hutchinson, Mark Lee, and Martin Rusch for numerous invaluable discussions. I am grateful to the Royal Society of London for financial support.
\end{acknowledgments}

\appendix

\section{Definition of dynamic terms with asymmetric summations} \label{sec:app:shifts}

In this Appendix we provide expressions for all dynamic terms in the theory. These expressions are obtained without symmetrizing with respect to the quasiparticle labels involved and are therefore appropriate for the new asymmetric summation technique introduced in Sec.~\ref{sec:convergence} as a means of substantially improving the numerical convergence. The results given here replace the equivalent ones obtained in Ref.~\cite{Morgan04}, to which they reduce in the case of an exact calculation or a symmetric summation. We will, however, make frequent reference to this earlier work for the definition of numerous matrix elements which appear unchanged in the formulae derived here.

Expressions for the fluctuations of all non-condensate mean-fields can be obtained by writing them in terms of time-dependent quasiparticle wavefunctions $u_i\rt$ and $v_i\rt$ and linearising these around their static values, $u_i\rt \rightarrow u_i(\bfr) + \delta u_i\rt$ etc. These quasiparticle fluctuations are found by introducing an expansion in the equilibrium basis of the form
\beq
\bpm \delta u_i\rt\\ \delta v_i\rt\epm = \sum_j X_{ij}(t)\bpm u_j(\bfr)\\ v_j(\bfr)\epm,
\label{app_Xqpexpansion}
\eeq
and obtaining the expansion coefficients $X_{ij}$ from the solution of a linearized form of the time-dependent BdG equations of Eq.~(\ref{uvt}) in the frequency domain. The results are given in Eq.~(A33) of Ref.~\cite{Morgan04} (see also Eqs.~(\ref{app_XcP})-(\ref{app_X0s}) below) and can be taken over unchanged to the present work (the formulae for $X_{ij}$ to not depend on whether we subsequently use a symmetric or asymmetric summation over the indices $i$ and $j$).

Following this procedure, the fluctuations in the non-condensate density and anomalous average are given by
\begin{align}
\delta\nt\rrw &= \sum_{ij \geq 0}  [N_i v_j^* u_i^* +  (N_i + 1)v_i^* u_j^*]X_{i,-j}(\omega) \nonumber\\ 
&+ \sum_{ij \geq 0} [N_i u_i v_j + (N_i + 1) u_j v_i] X_{i,-j}^*(-\omega)\nonumber\\ 
&+ \sum_{ij \geq 0} [N_i u_j u_i^* + (N_i + 1) v_i^* v_j] X_{ij}(\omega) \nonumber\\
&- \sum_{ij \geq 0} [N_i u_i u_j^* + (N_i + 1) v_j^* v_i] X_{ji}(\omega),
\label{app_dntrrw}
\end{align}
\begin{align}
\delta\mt^{\text{R}}\rrw &= \sum_{ij \geq 0}  [N_i v_j^* v_i^* +  (N_i + 1)v_i^* v_j^*]X_{i,-j}(\omega) \nonumber\\ 
&+ \sum_{ij \geq 0} [N_i u_i u_j + (N_i + 1) u_j u_i] X_{i,-j}^*(-\omega)\nonumber\\ 
&+ \sum_{ij \geq 0} [N_i u_j v_i^* + (N_i + 1) v_i^* u_j] X_{ij}(\omega) \nonumber\\
&- \sum_{ij \geq 0} [N_i u_i v_j^* + (N_i + 1) v_j^* u_i] X_{ji}(\omega)\nonumber\\
&+ N_0\frac{\Delta U_0}{U_0}\Phi_0 \delta\Phi\rw \delta(\bfr-\bfrp),
\label{app_dmtrrw}
\end{align}
where we have used the convention that the first wavefunction has the spatial argument $\bfr$ and the second $\bfrp$. The last line of Eq.~(\ref{app_dmtrrw}) is the contribution from the UV-renormalization. The summations in these equations are over all positive-norm modes and therefore include the corresponding zero-energy mode with a population factor $N_{0+} = 0$ (N.B. $N_{0+} \neq N_0$).

All dynamic terms required in the theory are projections of linear combinations of $\delta\nt\rrw$ and $\delta\mt^{\text{R}}\rrw$, usually, but not always, with $\bfr = \bfrp$. A generic dynamic term can therefore be written in the form
\begin{align}
\Delta_{p}^{(D)}(\omega)  = \sum_{ij \geq 0} f(N_i) \Biggl \{ &  \frac{C_{pij}^{(A)*}X_{i,-j}(\omega)}{2\sqrt{N_0}} + \frac{C_{pij}^{(B1)*}X_{i,-j}^*(-\omega)}{2\sqrt{N_0}} \nonumber\\
&+ \frac{C_{pji}^{(B2)*}X_{ij}(\omega)}{2\sqrt{N_0}} - \frac{C_{pij}^{(B2)*}X_{ji}(\omega)}{2\sqrt{N_0}} \Biggr \}\nonumber\\
& + \Delta_{p}^{\text{R}}(\omega),
\label{app_Deltap}
\end{align}
where the $C_{pij}$ coefficients are matrix elements involving integrals of quasiparticle wavefunctions whose definition depends on the particular combination of $\delta\nt\rrw$ and $\delta\mt^{\text{R}}\rrw$ appearing in the dynamic term in question. The population factor $f(N_i)$ is equal to $2N_i$ for finite temperature contributions and unity for zero temperature contributions, while the final term $\Delta_{p}^{\text{R}}(\omega)$ is the UV-renormalization (if any).

In an exact calculation where $\sum_{ij}$ ranges over all pairs of states, we can symmetrize the summands in the above equations with respect to $i$ and $j$. Eqs.~(\ref{app_dntrrw}) and (\ref{app_dmtrrw}) then reduce to Eqs.~(A26) and (A27) given in \cite{Morgan04}, and all expressions derived from them similarly reduce to the corresponding expressions given in Ref.~\cite{Morgan04}. In this case the Beliaev and Landau terms [respectively the first and second lines of Eq.~(\ref{app_Deltap})] acquire the familiar population factors of $1+N_i+N_j$ and $N_j-N_i$ respectively. However, any numerical calculation necessarily involves a finite basis and, as discussed in the main text, we should take the associated cutoff at a higher energy for the $j$ summation than for the $i$ summation to improve the convergence properties. Thus numerically, the summation is
\beq
\sum_{ij \geq 0} = \sum_{ij \geq 0} \Theta(E_1 - \epsilon_i)\Theta(E_2 - \epsilon_j),
\eeq
with $E_2 > E_1$ and $\Theta(x)$ the unit step function. This asymmetric summation requires the use of the expressions given in this appendix in place of those in Ref.~\cite{Morgan04}. It also has the unfortunate side-effect that for many quantities of interest, the matrix elements at zero-temperature differ slightly from their finite temperature values. This difference arises from the fact that the quasiparticle expansion of $\nt\rt$ has a different form at zero and finite temperature, cf. Eq.~(\ref{nt_qp}).

The coefficients $X_{ij}$ in Eq.~(\ref{app_Xqpexpansion}) contain two distinct contributions, one describing excitation of the thermal cloud by the condensate and the other its direct excitation by the external perturbation, as in Eq.~(A33) of \cite{Morgan04}, which we denote here by $X_{ij}^{(c)}$ and $X_{ij}^{(P)}$ respectively. In general, $X_{ij}^{(c)}$ depends on a sum over all excited condensate excitations, but if a single mode dominates the summation we can write $X_{ij}$ as
\beq
X_{ij}(\omega) = X_{ij}^{(c)}(p,\omega)b_p(\omega) + X_{ij}^{(P)}(\omega),
\label{app_XcP}
\eeq
where $b_p(\omega)$ is the condensate expansion coefficient of Eq.~(\ref{bpF}). For the case that neither of the subscripts $i$ or $j$ refers to the positive-norm, zero-energy mode ($i,j \neq 0+$), Eq.~(A33) of Ref.~\cite{Morgan04} gives for the condensate contribution
\begin{subequations}
\begin{align}
\frac{X_{ji}^{(c)}(p,\omega)}{\sqrt{N_0}} &= \frac{Y_{pij}^{(B2)}}{\hbar \omega - \epsilon_i + \epsilon_j}, \\
\frac{X_{i,-j}^{(c)}(p,\omega)}{\sqrt{N_0}} &= \frac{-Y_{pij}^{(A)}}{\hbar \omega + \epsilon_i + \epsilon_j},  \\
\frac{X_{i,-j}^{(c)*}(p,-\omega)}{\sqrt{N_0}} &= \frac{Y_{pij}^{(B1)}}{\hbar \omega - \epsilon_i - \epsilon_j}, 
\end{align}\label{app_XYps}
\end{subequations}
while the contribution from the perturbation is
\begin{subequations}
\begin{align}
X_{ji}^{(P)}(\omega) &= \frac{P_{ij}^L(\omega)}{\hbar \omega - \epsilon_i + \epsilon_j}, \\
X_{i,-j}^{(P)}(\omega) &= \frac{-P_{ij}^{B*}(-\omega)}{\hbar \omega + \epsilon_i + \epsilon_j},  \\
X_{i,-j}^{(P)*}(-\omega) &= \frac{P_{ij}^B(\omega)}{\hbar \omega - \epsilon_i - \epsilon_j}. 
\end{align}\label{app_XPs}
\end{subequations}
The matrix elements $Y_{pij}$, $P_{ij}^B(\omega)$ and $P_{ij}^L(\omega)$ in these equations are defined in Eqs.~(106), (113) and (114) of \cite{Morgan04}. Their calculation involves integrals of three quasiparticle wavefunctions and the condensate (the $Y_{pij}$) or two quasiparticle wavefunctions and the external perturbation [$P_{ij}^B(\omega)$ and $P_{ij}^L(\omega)$].

If, either of the subscripts $i$ or $j$ refers to the zero-mode, however,
the formulae are different. In this case all the coefficients
$X_{ij}^{(P)}(\omega)$ are zero and the $X_{ij}^{(c)}(p,\omega)$ are given by (cf. Eqs.~(A35)-(A37) of \cite{Morgan04})
\begin{subequations}
\begin{gather}
X_{i,0+}^{(c)}(p,\omega) = X_{i,0-}^{(c)}(p,\omega) = X_{0+,-i}^{(c)}(p,\omega) = -W_{ip},\\
X_{0+,i}^{(c)}(p,\omega) = -X_{i,0+}^{(c)*}(p,-\omega) = U_{ip},\\ X_{i,0-}^{(c)*}(p,-\omega) = -V_{ip},\quad X_{0+,0+}^{(c)} = 0,
\end{gather}\label{app_X0s}
\end{subequations}
where $U_{ij}$, $V_{ij}$ and $W_{ij}$ involve integrals of products of two quasiparticle wavefunctions and are defined in Eqs.~(99)-(101) of \cite{Morgan04}.

We deal with the different expressions for these two cases by restricting the summation in Eq.~(\ref{app_Deltap}) to positive energy modes and calculating the zero-mode contribution separately. Our final expression for a generic dynamic term is therefore
\beq
\Delta_{p}^{(D)}(\omega)  = \Delta_{pp}^{(c)}(\omega)b_p({\omega}) + \Delta_{p}^{(P)}(\omega),
\label{app_DeltapcPb}
\eeq
where the condensate contribution $\Delta_{pp}^{(c)}(\omega)$ is further subdivided as
\beq
\Delta_{pp}^{(c)}(\omega)  = \Delta_{pp}^{(c+)}(\omega) +  \Delta_{0}(p) + \Delta^{\text{R}}(p).
\label{app_Deltacpp}
\eeq
Here $\Delta_{pp}^{(c+)}(\omega)$ and $\Delta_{p}^{(P)}(\omega)$ are calculated using Eq.~(\ref{app_Deltap}) with $\sum_{ij\geq 0} \rightarrow \sum_{ij > 0}$ and $X_{ij}(\omega) \rightarrow X_{ij}^{(c)}(p,\omega)$ or $X_{ij}^{(P)}(\omega)$ from Eqs.~(\ref{app_XYps})-(\ref{app_XPs}) respectively, while $\Delta_{0}(p)$ is the zero-mode contribution defined by
\begin{align}
\Delta_{0}(p) = 
 &-\frac{1}{\sqrt{N_0}} \sum_{i>0} \Bigl [ C_{p0i}^{(B2)*}W_{ip} N_i + C_{pi0}^{(A)*}W_{pi}(N_i+1) \Bigr ] \nonumber \\
 &-\frac{1}{\sqrt{N_0}} \sum_{i>0} \Bigl [ C_{pi0}^{(B2)*}U_{ip} N_i + C_{pi0}^{(B1)*}V_{ip}(N_i+1)\Bigr ].
\label{Deltap0}
\end{align}
$\Delta^{\text{R}}(p)$ is the UV-renormalization (if any), obtained from the last line of Eq.~(\ref{app_Deltap}) via $\Delta_{p}^{\text{R}}(\omega) = \Delta^{\text{R}}(p)b_p(\omega)$.

\subsection{Special cases}

In this paper, we are interested in the quantities $\Delta E_{pp}^{(D)}(\omega)$ and $\Delta P_{p0}^{(D)}(\omega)$, defined respectively as the condensate and perturbation parts of the particular projection of $\delta \nt$ and $\delta \mt^{\text{R}}$ given in Eq.~(A28) of \cite{Morgan04}. Writing $\Delta E_{pp}^{(D)}(\omega)$ in the form of Eq.~(\ref{app_Deltacpp}) we have
\beq
\Delta E_{pp}^{(D)}(\omega) = \Delta E_{pp}^{(+)}(\omega) +  \Delta E_{0}(p) + \Delta E^{\text{R}}(p).
\label{app_DEppD}
\eeq 
Comparing with Eq.~(\ref{app_Deltap}), we find that for $\Delta E_{pp}^{(+)}(\omega)$, $\Delta E_{0}(p)$ and $\Delta P_{p0}^{(D)}(\omega)$, the $C_{pij}$ coefficients are given at finite temperature ($f(N_i) = 2N_i$) by
\beq
C_{pij}^{(A)} = Y_{pij}^{(A)}, \quad 
C_{pij}^{(B1)} = Y_{pij}^{(B1)}, \quad
C_{pij}^{(B2)} = Y_{pij}^{(B2)},
\eeq
where the $Y_{pij}$ coefficients are the same as those in Eq.~(\ref{app_XYps}) and are defined in Eq.~(106) of \cite{Morgan04}. At zero temperature ($f(N_i) = 1$) these coefficients are given instead by
\begin{subequations}
\begin{align}
C_{pij}^{(A)} &= Y_{pij}^{(A)} + dA_{pij}, \\
C_{pij}^{(B1)} &=  Y_{pij}^{(B1)} + dB1_{pij}, \\
C_{pij}^{(B2)} &=  Y_{pij}^{(B2)} + dB2_{pij},
\end{align}
\end{subequations}
where the quantities $dA_{pij}$ etc are the changes to the integrals $A_{pij}$ etc defined in Eqs.~(107)-(109) of \cite{Morgan04} arising from the asymmetry in the zero-temperature part of $\delta\nt\rw$. They are given by
\begin{subequations}
\begin{align}
dA_{pij} &= 2\sqrt{N_{0}} U_{0} \! \intdr (u_p\Phi_0^* + v_p\Phi_0)(v_i u_j - u_i v_j), \\
dB1_{pij} &= 2\sqrt{N_{0}} U_{0} \! \intdr (u_p\Phi_0^* + v_p \Phi_0)(v_i u_j - u_i v_j)^*, \\
dB2_{pij} &= 2\sqrt{N_{0}} U_{0} \! \intdr (u_p \Phi_0^* + v_p \Phi_0)(v_i^* v_j - u_i^* u_j).
\end{align}
\label{app_dABB}
\end{subequations}
The UV-renormalization $\Delta E^{\text{R}}(p)$ is given by Eq.~(105) in \cite{Morgan04} and involves the quantity $\Delta U_0$ of Eq.~(\ref{DeltaU0}). For numerical consistency, we calculate $\Delta U_0$ using an integration up to $E_{cut}$ and only take the integral to infinity if we also include a semi-classical approximation for higher energy states.

To calculate the condensate number fluctuations $\delta N_0(\omega) = -\intdr \delta \nt \rw$ (which are only non-zero for modes with $m_p = p_p = 0$), we have at finite temperature ($f(N_i) = 2N_i$)
\beq
C_{pij}^{(A)} = C_{pij}^{(B1)*} = -\sqrt{N_0}J_{ij}, \quad
C_{pij}^{(B2)} = -\sqrt{N_0}I_{ij},
\eeq
where $I_{ij}$ and $J_{ij}$ are defined in Eqs.~(110)-(111) of \cite{Morgan04}.
At zero temperature ($f(N_i) = 1$), the coefficients $C_{pij}^{(A)}$ and $C_{pij}^{(B1)}$ are unchanged, while $C_{pij}^{(B2)}$ becomes
\beq
C_{pij}^{(B2)} = -2\sqrt{N_0}V_{ij},
\eeq
with $V_{ij}$ as in Eq.~(100) of \cite{Morgan04}. In this case there is no zero-mode contribution because the condensate is orthogonal to states of finite energy, and there is also no UV-renormalization.

\subsection{Condensate density fluctuations} \label{sec:app:condDensflucs}

Experiments generally measure moments of the condensate density fluctuations given in Eq.~(\ref{dnc_GGPE}). If only the single mode `p' is excited these fluctuations have the spatial form $u_p(\bfr)\ccond +v_p(\bfr)\cond$. Since the density fluctuations are real in the time-domain, we calculate their projection onto the real part of this function, $f_p(\bfr) = \mbox{Re}[u_p^*(\bfr)\cond +v_p^*(\bfr)\ccond]$. Defining the integral
\beq
d_p = \intdr f_p(\bfr)[u_p(\bfr)\ccond +v_p(\bfr)\cond],
\eeq
we therefore calculate the quantities
\begin{align}
\mathcal{G}_p^{\delta n_c}(\omega) &\mbox{ or } \mathcal{R}_p^{\delta n_c}(\omega) \nonumber \\
&= \frac{1}{N_0 d_p P_{p0}(\omega)}\intdr f_p(\bfr)\delta  n_c \rw  \label{app_GRdN0}\\
&= \frac{\mbox{Re}(c_p)}{N_0 U_0 d_p} \left [\frac{\delta N_0(\omega)}{N_0P_{p0}(\omega)} + \frac{2b_{N_0}(\omega)}{P_{p0}(\omega)} \right] + \frac{b_p(\omega)}{P_{p0}(\omega)},\nonumber
\end{align}
where $c_p$ is defined by the equilibrium limit of Eq.~(\ref{ci}),  $b_{N_0}(\omega)$ is the coefficient defined in Eq.~(81) of Ref.~\cite{Morgan04} and we obtain either $\mathcal{G}_p^{\delta n_c}(\omega)$ or  $\mathcal{R}_p^{\delta n_c}(\omega)$ on the left-hand-side depending on whether or not we exclude direct driving of the non-condensate in evaluating the expressions on the right-hand-side. If either $m_p \neq 0$ or $p_p \neq 0$, then $\delta N_0(\omega) = b_{N_0}(\omega) = 0$ and $\mathcal{G}_p^{\delta n_c}(\omega)$ and $\mathcal{R}_p^{\delta n_c}(\omega)$ reduce to the resolvents $\mathcal{G}_p(\omega)$ and $\mathcal{R}_p(\omega)$ defined in the main text in Eqs.~(\ref{G}) and (\ref{R}) respectively. We see therefore that these quantities are directly proportional to the condensate density fluctuations measured in experiments.

\subsection{Non-condensate density fluctuations} \label{sec:app:ncfluctuations}

To calculate the relative phase of condensate--non-condensate oscillations, or simply to compare the relative sizes of condensate and non-condensate fluctuations, we also need to calculate the projection of $\delta\nt\rw$ onto the function $f_p(\bfr)$ defined above. For direct comparison with the resolvents $\mathcal{G}_p^{\delta n_c}(\omega)$ or  $\mathcal{R}_p^{\delta n_c}(\omega)$, we actually calculate
\beq
\mathcal{G}_p^{\delta \nt}(\omega) \mbox{ or } \mathcal{R}_p^{\delta \nt}(\omega) = 
\frac{1}{N_0 d_p P_{p0}(\omega)}\intdr f_p(\bfr) \delta  \nt\rw,
\label{app_GRdnt}
\eeq
where we obtain either $\mathcal{G}_p^{\delta\nt}(\omega)$ or  $\mathcal{R}_p^{\delta\nt}(\omega)$ if we exclude or include direct driving of the non-condensate respectively. Thus the total density fluctuation projections are proportional to $\mathcal{G}_p^{\delta n_c}(\omega) + \mathcal{G}_p^{\delta \nt}(\omega)$ or $\mathcal{R}_p^{\delta n_c}(\omega) + \mathcal{R}_p^{\delta \nt}(\omega)$. The phase of the non-condensate oscillations relative to those of the condensate can be found as a function of frequency from the argument of the complex quantities $[\mathcal{G}_p^{\delta \nt}(\omega)]^{*} \times\mathcal{G}_p^{\delta n_c}(\omega)$ or $[\mathcal{R}_p^{\delta \nt}(\omega)]^{*} \times\mathcal{R}_p^{\delta n_c}(\omega)$.

The projection $\intdr f_p(\bfr) \delta  \nt\rw$ can be found from Eq.~(\ref{app_dntrrw}) and has the generic form given in Eq.~(\ref{app_Deltap}). For numerical convenience we actually calculate the quantity $2U_0\intdr [u_p^* \Phi_0 + v_p^* \Phi_0^*] \delta  \nt\rw$, for which the various $C_{pij}$ coefficients are given at finite temperature ($f(N_i) = 2N_i$) by
\begin{subequations}
\begin{align}
C_{pij}^{(A)} &= 2\sqrt{N_{0}} U_{0} \! \intdr (u_p\Phi_0^* + v_p\Phi_0)(u_i v_j + v_i u_j), \\
C_{pij}^{(B1)} &= 2\sqrt{N_{0}} U_{0} \! \intdr (u_p\Phi_0^* + v_p \Phi_0)(u_i v_j + v_i u_j)^*, \\
C_{pij}^{(B2)} &= 2\sqrt{N_{0}} U_{0} \! \intdr (u_p \Phi_0^* + v_p \Phi_0)(u_i^* u_j + v_i^* v_j).
\end{align}
\end{subequations}
For the zero temperature contribution ($f(N_i) = 1$) these coefficients are modified according to
\begin{subequations}
\begin{align}
C_{pij}^{(A)} &\rightarrow C_{pij}^{(A)} + dA_{pij}, \\
C_{pij}^{(B1)} &\rightarrow  C_{pij}^{(B1)} + dB1_{pij}, \\
C_{pij}^{(B2)} &\rightarrow  C_{pij}^{(B2)} + dB2_{pij},
\end{align}
\end{subequations}
with $dA_{pij}$ etc as in Eq.~(\ref{app_dABB}). No UV renormalization is required in this case so $\Delta^{\text{R}}_{\delta n_c}(p) = 0$.

\section{semi-classical approximation} \label{sec:app:semi-classical}

The calculation of both the static and dynamic terms may be supplemented with a semi-classical approximation for the quasiparticle modes above the numerical cutoff energy to improve convergence. This procedure is straightforward to implement for the static terms, and is very accurate even for quite low values of the cutoff. Unfortunately, the situation is much more complicated for the dynamic terms, however, and additional approximations are required to reduce the expressions to a manageable form \cite{Giorgini00}. This limits the practical accuracy of the method in this case and we have found numerically that the approximation we implement for the dynamic terms only becomes adequate at quite high energies and must be used in conjunction with a large numerical basis (see Sec.~\ref{sec:convergence} and Fig.~\ref{fig:Econv201}).

The semi-classical approximation is described in detail in Refs.~\cite{Giorgini00,Reidl99,Giorgini97}. The essence of the method is that the quantum numbers `p' labelling a particular quasiparticle wavefunction $u_p(\bfr)$ become a momentum label $\bfp$ so that the new wavefunction is defined in a single-particle phase space $(\bfp,\bfr)$ by
\begin{align}
u_p(\bfr) \rightarrow u(\bfp,\bfr) &= \bar{u}(\bfp,\bfr)\frac{e^{i\bfp.\bfr/\hbar}}{\sqrt{V}},\\
v_p(\bfr) \rightarrow v(\bfp,\bfr) &= \bar{v}(\bfp,\bfr)\frac{e^{i\bfp.\bfr/\hbar}}{\sqrt{V}},
\end{align}
where $V$ is some suitable volume. The exponential factor contains the `fast' dependence of the functions on the coordinates, while $\bar{u}(\bfp,\bfr)$ and $\bar{v}(\bfp,\bfr)$ only vary on a scale set by the size of the condensate and the chemical potential. The semi-classical solution to the BdG equations is then given by
\begin{gather}
\bar{u}^2(\bfp,\bfr) = 1+\bar{v}^2(\bfp,\bfr) = \frac{\epsilon_{\scriptscriptstyle HF}(\bfp,\bfr) + \epr}{2\epr},\\
\bar{u}(\bfp,\bfr)\bar{v}(\bfp,\bfr) = -\frac{n_0(\bfr)U_0}{2\epr},
\end{gather}
where $n_0(\bfr) = N_0|\Phi_0(\bfr)|^2$ is the local condensate density and $\epsilon(\bfp,\bfr)$ and $\epsilon_{\scriptscriptstyle HF}(\bfp,\bfr)$ are the local quasiparticle and Hartree-Fock energies respectively, defined by
\begin{gather}
\epsilon^2(\bfp,\bfr) = \epsilon_{\scriptscriptstyle HF}^2(\bfp,\bfr)-(n_0(\bfr)U_0)^2,\\
\epsilon_{\scriptscriptstyle HF}(\bfp,\bfr) = \frac{\bfp^2}{2m} + V_{\text{trap}}(\bfr) - \lambda_0 + 2n_0(\bfr)U_0.
\end{gather}
Summations over quasiparticle states are replaced by integrations over momenta via
\beq
\frac{1}{V}\sum_i \rightarrow \int \!\! \frac{d^{3}\bfp}{(2\pi\hbar)^3}.
\eeq

We therefore obtain the following semi-classical approximations to the equilibrium non-condensate mean-fields $\tilde{n}(\bfr)$ and $\tilde{m}^{\text{R}}(\bfr)$ needed in the static terms
\beq
\tilde{n}_{sc}(\bfr) = \int \!\! \frac{d^{3}\bfp}{(2\pi\hbar)^3} \bigl [ (\bar{u}^2 + \bar{v}^2)N(\epsilon) + \bar{v}^2 \bigr ]\Theta(\epsilon-E_{cut}),
\label{app_ntsc}
\eeq
\begin{align}
\tilde{m}_{sc}^{\text{R}}(\bfr) = &-\int \!\! \frac{d^{3}\bfp}{(2\pi\hbar)^3} \frac{n_0U_0[2N(\epsilon) + 1]}{2\epsilon}\,\Theta(\epsilon-E_{cut}) \nonumber \\
&+ \int \frac{d^{3}\bfp}{(2\pi\hbar)^3} \frac{n_0U_0}{2(\bfp^2/2m)}.
\label{app_mtsc}
\end{align}
The last line of the expression for $\tilde{m}_{sc}^{\text{R}}(\bfr)$ contains the full UV renormalization of the anomalous average (necessary for convergence at zero temperature) and is integrated over all energies, not just those above $E_{cut}$. In practice, we divide this integral into two regions, corresponding to energies above and below the cutoff. The contribution from energies below the cutoff is included in the exact basis calculation of $\tilde{m}^{\text{R}}(\bfr)$ so that this is well-behaved as the cutoff increases, while the part above the cutoff is calculated along with the remaining semi-classical correction in Eq.~(\ref{app_mtsc}).

The integrals over the direction of the momentum in Eqs.~(\ref{app_ntsc}) and (\ref{app_mtsc}) are trivial so the semi-classical approximation for the static terms reduces to the calculation of one-dimensional integrals over energy for each spatial grid point. In this case, the approximation rapidly becomes highly accurate, so if only static terms are required in the theory a small numerical basis is sufficient. 

The problem of how to apply the semi-classical approximation to dynamic terms of the form of Eq.~(\ref{dD}) has been studied in detail by Giorgini \cite{Giorgini00}. In a tour-de-force calculation, he combined the semi-classical treatment of the excitations with a Thomas-Fermi approximation for the condensate (necessary for consistency at low energy \cite{Reidl99}) to obtain analytical predictions for energy shifts and decay rates for trapped gases in the thermodynamic limit. Unfortunately the method is limited to modes with constant Laplacian because of the nature of additional approximations which have to be made to deal with the Landau terms. Although these modes include those studied to date, we require a method which can be used for finite systems outside the Thomas-Fermi regime. This can be obtained by adapting the method described by Giorgini \cite{Giorgini00} to the current problem. 

We start by writing the semi-classical approximation to the dynamic terms in the form (we focus on $\Delta E_p^{(D)}$ in the following)
\begin{align}
\Delta E_{sc}^{(D)}(p) = \frac{1}{(2\pi\hbar)^6}\iiiint \!\!d^{3}\bfr \, d^{3}\bfs \, d^{3}\bfp_i \, d^{3}\bfq \, e^{-i\bfq.\bfs}& \nonumber\\
g_p^*(\bfr)K_{ij}(\bfr,\bfr+\bfs,\omega)g_p(\bfr+\bfs)&
\end{align}
where $\bfq = \bfp_i-\bfp_j$, $\bfs = \bfrp-\bfr$ and $g_p(\bfr)$ is a smooth function of $\bfr$ dependent only on $\cond$, $u_p(\bfr)$ and $v_p(\bfr)$ (these are the full numerical solutions and do not involve any semi-classical approximation). The kernel $K_{ij}(\bfr,\bfr+\bfs,\omega)$ contains all the slow dependence of the intermediate states, i.e. the energy denominators, population factors and the slowly-varying part of the quasiparticle wave functions. The exponential factor $e^{-i\bfq.\bfs}$ contains the rapidly varying part of the quasiparticle wave functions and acts like a delta function in both position and momentum, so that an expansion in powers of $\bfs$ is appropriate. The leading order corresponds to setting $\bfs = 0$ in the kernel and integration over $\bfs$ in the exponential then leaves a delta function of $\bfq$. The double integrals over position and momentum therefore collapse into single integrals as for the static shifts, and we obtain
\beq
\Delta E_{sc}^{(D)}(p) = \frac{1}{(2\pi\hbar)^3}\iint \!\!d^{3}\bfr \, d^{3}\bfp_i \, 
|g_p(\bfr)|^2 K_{ii}(\bfr,\bfr,\omega).
\eeq

The main difficulty with applying this approach is the pole in the kernel for the Landau terms where $K_{ij}^{L} \sim (N_j-N_i)/(\omega -\epsilon_i+\epsilon_j)$. This invalidates the assumption that the kernel is slowly varying and restricts the analytical calculation to modes with constant Laplacian \cite{Giorgini00}. However, we are only interested in the semi-classical approximation above a large cutoff energy, and we therefore expect that the result will be approximately independent of frequency in the range of interest. We therefore set $\omega = 0$ in the kernels, which allows us to replace the badly-behaved factor $(N_j-N_i)/(\omega -\epsilon_i+\epsilon_j)$ with the derivative $-(dN/dE)|_{\epsilon_i}$. The final form of the semi-classical approximation that we actually use for $\Delta E_{p}^{(D)}$ is therefore
\begin{align}
\Delta E_{sc}^{(D)}(p) = \iint \frac{d^{3}\bfr \, d^{3}\bfp_i}{(2\pi\hbar)^3}\Theta(\epsilon-E_{cut})\, \left\{\right. \left(\frac{dN}{d\epsilon}\right)[Y_{pii}^{(B2)}(\bfr)]^2 & \nonumber \\
- \frac{(1+2N)([Y_{pii}^{(A)}(\bfr)]^2 + [Y_{pii}^{(B1)}(\bfr)]^2)}{4\epr} \left. \right\},&
\label{app_dEsc}
\end{align}
where the coefficients $\bar{Y}_{pii}(\bfr)$ are the integrands of the coefficients $Y_{pij}$ appearing in Eq.~(\ref{app_XYps}) and defined in Eq.~(106) of \cite{Morgan04} but with the exact quasiparticle wavefunctions replaced by their slowly varying counterparts, i.e. $u_i({\bfr}) \rightarrow \bar{u}(\bfp_i,\bfr)$ etc. We have found numerically that the approximation of Eq.~(\ref{app_dEsc}) is much less accurate than the semi-classical results for the static terms but  becomes adequate when a high cutoff energy is used. The effect of the semi-classical approximation on our numerical convergence can be seen in Fig.~\ref{fig:Econv201}.

We have not used a semi-classical approximation for the dynamic term $\Delta P_{p0}^{(D)}(\omega)$ because the long length scale of the external potential invalidates the approach. This is not particularly serious, however, as there is no delicate cancellation of terms in the numerator of the response function as there is in the denominator so the accuracy requirement is greatly reduced. In addition, the use of the asymmetric summation technique described in Sec.~\ref{sec:convergence} removes any need for a semi-classical approximation for this term.


\begin{thebibliography}{99}
\bibitem{dropDN0} For simplicity, we have neglected the contribution to this equation from statistical fluctuations in the equilibrium condensate population (introduced in \cite{Morgan04}) as these have a negligible effect on our numerical results.
\bibitem{cite:groundstate} Although our numerical results in this paper apply to the case that the equilibrium condensate wavefunction is the ground state of the GPE, the theoretical formalism developed in \cite{Morgan04} is more general and also applies to excited state solutions, such as those containing vortices.
\bibitem{cite:table} The results in Table~\ref{table:energies} are calculated using a symmetric summation for dynamic terms because we wanted to present meaningful results for individual contributions to the self-energy and not just for the total. This requires the use of the semi-classical approximation to ensure that each term has individually converged and we have found numerically that this approximation does not work reliably for the new asymmetric technique. Investigating this issue will be a subject for future work, but fortunately the asymmetric summation approach gives such good convergence for the total self-energy that the semi-classical approximation is not required for the results presented elsewhere in the paper.
\bibitem{errorbars} The experimental results were extracted from the postscript of Fig.~1 in Ref.~\cite{Jin97} because the raw data no longer exists. The size of the circles in our Fig.~\ref{fig:TNN0vst} approximately represents the uncertainty in this operation.
\end{thebibliography}
\end{document}